\documentclass[%
reprint,
nofootinbib,
noeprint
 amsmath,amssymb,
 aps,
pra,
]{revtex4-2}

\usepackage{graphicx}
\usepackage{dcolumn}
\usepackage{bm}
\usepackage{bbm}
\usepackage{gensymb}
\usepackage{physics}
\usepackage{setspace}
\usepackage{orcidlink}
\usepackage{url}
\usepackage{hyperref}
\hypersetup{
  colorlinks=true,
  allcolors=blue
}

\begin{document}

\title{Chirality-bolstered quantum Zeno effect enhances radical pair-based magnetoreception}

\author{Luke D.\ Smith$^{1, 2}$ \orcidlink{0000-0002-6255-2252}}

\author{Sukesh\ Tallapudi$^{3}$ \orcidlink{0009-0009-4412-6223}}

\author{Matt C.\ J.\ Denton$^{1, 2}$ \orcidlink{0009-0009-0246-3387}}

\author{Daniel R.\ Kattnig$^{1, 2, *}$ \orcidlink{0000-0003-4236-2627}}
\affiliation{$^{1}$Living Systems Institute, University of Exeter, Stocker Road, Exeter, Devon, EX4 4QD, United Kingdom.
}
\affiliation{$^{2}$Department of Physics, University of Exeter, Stocker Road, Exeter, Devon, EX4 4QL, United Kingdom.}
\affiliation{$^{3}$Indian Institute of Science, Bangalore 560012, India.}
\affiliation{$^{*}$d.r.kattnig@exeter.ac.uk}

\begin{abstract}
Radical pairs in the flavoprotein cryptochrome are central to various magnetically sensitive biological processes, including the proposed mechanism of avian magnetoreception. Cryptochrome’s molecular chirality has been hypothesized to enhance magnetic field effects via the chirality-induced spin selectivity (CISS) effect, yet the mechanism underlying this enhancement remains unresolved. In this work, we systematically investigate the impact of CISS on the directional magnetic sensitivity of prototypical radical pair reactions, analyzing two distinct models--one generating spin polarization and, for the first time, one generating coherence. We find that CISS-induced spin polarization significantly enhances magnetic sensitivity by introducing triplet character into the initial state and reinforcing the quantum Zeno effect, paralleling enhancements observed in triplet-born radical pairs subject to strongly asymmetric recombination. In contrast, CISS-generated spin coherence does not provide a significant improvement in sensitivity. These findings indicate that CISS is not itself a universal enhancer of sensitivity or coherence in radical-pair reactions, and its influence must be evaluated case by case, particularly in relation to the quantum Zeno effect. Additionally, we provide a unified interpolation scheme for modeling CISS-influenced initial states and recombination dynamics, encompassing the principal models currently discussed in the literature for singlet and triplet precursors.
\end{abstract}

\maketitle

\section*{Introduction}

The radical pair mechanism (RPM) provides the most compelling explanation for magnetic field effects (MFEs) in biology, including the proposed visual magnetoreception of migratory birds \cite{Wiltschko2019, Mouritsen2018, Johnsen2008}. The RPM describes how spin dynamics within radical pairs, appearing as short-lived intermediates in electron transfer reactions, acquire magnetic field sensitivity as a result of their coherent spin dynamics \cite{Hore2016}. A particularly actively researched field in this context is (avian) magnetoreception, which is considered a prime example of quantum biology \cite{alvarezQuantumPhenomenaBiological2024, Kim2021}. The flavoprotein cryptochrome is the leading candidate for hosting this sensory function \cite{Ritz2000}.

In cryptochrome, two radical pairs are widely considered. The FAD$^{\bullet -}$/W$_{\mathrm{C}}^{\bullet+}$ radical pair is generated via a photo-induced electron transfer from a tryptophan residue ($\mathrm{W_{C}}$) to a flavin adenine dinucleotide (FAD) cofactor \cite{ramsayCryptochromeMagnetoreceptionTime2024, frederiksenMutationalStudyTryptophan2023a, Wong2021,Kattnig2016, nohrExtendedElectronTransferAnimal2016, mullerDiscoveryFunctionalAnalysis2015, liedvogelChemicalMagnetoreceptionBird2007, giovaniLightinducedElectronTransfer2003}. The second, FADH$^{\bullet}$/O$^{\bullet-}_2$, forms in the dark during flavin reoxidation with molecular oxygen \cite{deviersAvianCryptochrome42024a, salernoLongTimeOxygenSuperoxide2023, arthautBluelightInducedAccumulation2017, vanwilderenKineticStudiesOxidation2015, Muller2011, Ritz2009, masseyActivationMolecularOxygen1994}. While the former is singlet-born and has demonstrated MFEs under moderate magnetic fields \textit{in vitro} \cite{Xu2021, Kerpal2019, Maeda2008, Timmel1998}, its weak-field magnetosensitivity is attenuated by interradical interactions \cite{Babcock2020, Efimova2008, ODea2005}, spin relaxation \cite{gruningEffectSpinRelaxation2024, Worster2016, Kattnig2016a}, and symmetrically distributed but misaligned hyperfine couplings \cite{Gruning2022, Smith2022, Atkins2019}. The latter, triplet-born FADH$^{\bullet}$/O$^{\bullet-}_2$, benefits from an asymmetric distribution of hyperfine couplings over the two radicals \cite{smithOptimalityRadicalpairQuantum2024, Procopio2020, Lee2014}, but suffers from rapid spin relaxation via the spin-rotational mechanism when tumbling freely \cite{Player2019, karogodinaKineticMagneticfieldEffect2011, Hogben2009}, or strong interradical coupling when immobilized \cite{deviersAvianCryptochrome42024a, salernoLongTimeOxygenSuperoxide2023, mondalTheoreticalInsightsFormation2019}. For both systems, the mentioned adverse effects raise doubts about the viability of sensing the direction of the weak geomagnetic field. This has led to an active search for enhancing mechanisms over the past decade, which has hinted at remarkable possibilities.

Several enhancements to RPM have been proposed to counter the sensitivity gap: J/D cancellation \cite{Efimova2008}, three-radical mechanisms \cite{Babcock2020, Keens2018}, radical scavenging \cite{Ramsay2022, Deviers2022, Babcock2021, Kattnig2017}, relaxation-assisted mechanisms \cite{luoSensitivityEnhancementRadicalpair2024a,Kattnig2017} and both diffusive and driven-radical motion \cite{Ramsay2023, Smith2022}. Although most of these extensions are effective in enhancing the magnetosensitivity of cryptochrome, they require additional features, the viability of which in real systems is yet to be determined. More recently, enhancements through asymmetrical kinetics and the quantum Zeno effect \cite{burgarthQuantumZenoDynamics2020, Dellis2012, berdinskiiChemicalZenoEffect2008, itanoQuantumZenoEffect1990, misraZenosParadoxQuantum1977} have shown promise, particularly in tightly bound triplet-born radical pairs such as FADH$^{\bullet}$/O$^{\bullet-}_2$ \cite{dentonMagnetosensitivityTightlyBound2024}.

Among the various spin-dependent phenomena in chemistry, the chirality-induced spin selectivity (CISS) effect has recently gained significant attention for its proposed role in polarizing electron spins in both artificial and biological systems \cite{chiesaChiralityInducedSpinSelectivity2023,aielloChiralityBasedQuantumLeap2022, abendrothSpinSelectivityPhotoinduced2019, bloomChiralityControlElectron2017, michaeliElectronsSpinMolecular2016, wasielewskiEnergyChargeSpin2006}. Generally, CISS describes the coupling between chirality and electron spin in electron displacement currents and transmission, such as in electron transfer reactions, which can induce spin coherence \cite{fayChiralityInducedSpinCoherence2021}, and spin polarization \cite{faySpinSelectiveCharge2023, vittmannSpinDependentMomentumConservation2023, vittmannInterfaceInducedConservationMomentum2022, fayOriginChiralityInduced2021, Luo2021, chiesaAssessingNatureChiralInduced2021}. A comprehensive review of CISS and its theoretical modeling can be found in \cite{bloomChiralInducedSpin2024}. However, the theoretical basis for the CISS effect remains incomplete. This notwithstanding, it has been suggested that CISS may influence radical pair dynamics, for example, through spin polarization during formation or recombination \cite{Luo2021}.

Observing that proteins, such as cryptochrome, are inherently chiral raises an intriguing question: Could the CISS effect play a role in quantum biology, especially in cryptochrome \cite{alvarezQuantumPhenomenaBiological2024, Kim2021}? In fact, preliminary explorations appear to support this idea. Studies on FAD$^{\bullet -}$/W$_{\mathrm{C}}^{\bullet+}$ radical pairs have demonstrated promising sensitivity and coherence enhancements \cite{Poonia2023, Poonia2022} using a phenomenological model that assumes spin polarization during the formation and recombination of the radical pair \cite{Luo2021}. However, the model used is not supported by microscopic derivation and is not universal. Moreover, the predicted enhancements are, as of now, not supported by direct experimental validation, for which even the existence of the effect in cryptochrome has yet to be confirmed. Although often attributed to spin-orbit coupling, research suggests that chirality and spin-orbit coupling alone induce spin coherence but not spin polarization \cite{fayChiralityInducedSpinCoherence2021}, which also requires electron hopping and electronic exchange interactions \cite{fayOriginChiralityInduced2021}. The effects of CISS-generated spin coherence alone--without spin polarization--on the magnetosensitivity of radical pairs in cryptochrome have not yet been studied. Furthermore, observing that CISS enhancements suspiciously appear when strongly asymetric reaction kinetics apply raises the question whether the observed enhancements might instead arise from the quantum Zeno effect.

In this work, we aim to clarify the role of CISS in enhancing the magnetic-field sensitivity of cryptochrome radical pairs. Specifically, we ask: Can CISS-induced spin polarization or coherence enhance magnetosensitivity in cryptochrome? How do these effects interact with the quantum Zeno mechanism? Are enhancements feasible without spin polarization? To address these questions, we analyze both spin polarization and coherence-generating CISS models in two cryptochrome radical pairs (FAD$^{\bullet -}$/W$_{\mathrm{C}}^{\bullet+}$ and FADH$^{\bullet}$/O$^{\bullet-}_2$), using a theoretical framework based on the Nakajima–Zwanzig projection operator formalism \cite{fayElectronSpinRelaxation2019}. We derive general expressions for CISS-modified recombination operators and investigate sensitivity enhancements across different initial conditions and kinetic regimes. Our results clarify the microscopic basis and limits of CISS in quantum biological systems, with implications for magnetoreception and spin-based bioelectronics.

\begin{figure*}
     \centering
     \includegraphics[width=.9\linewidth]{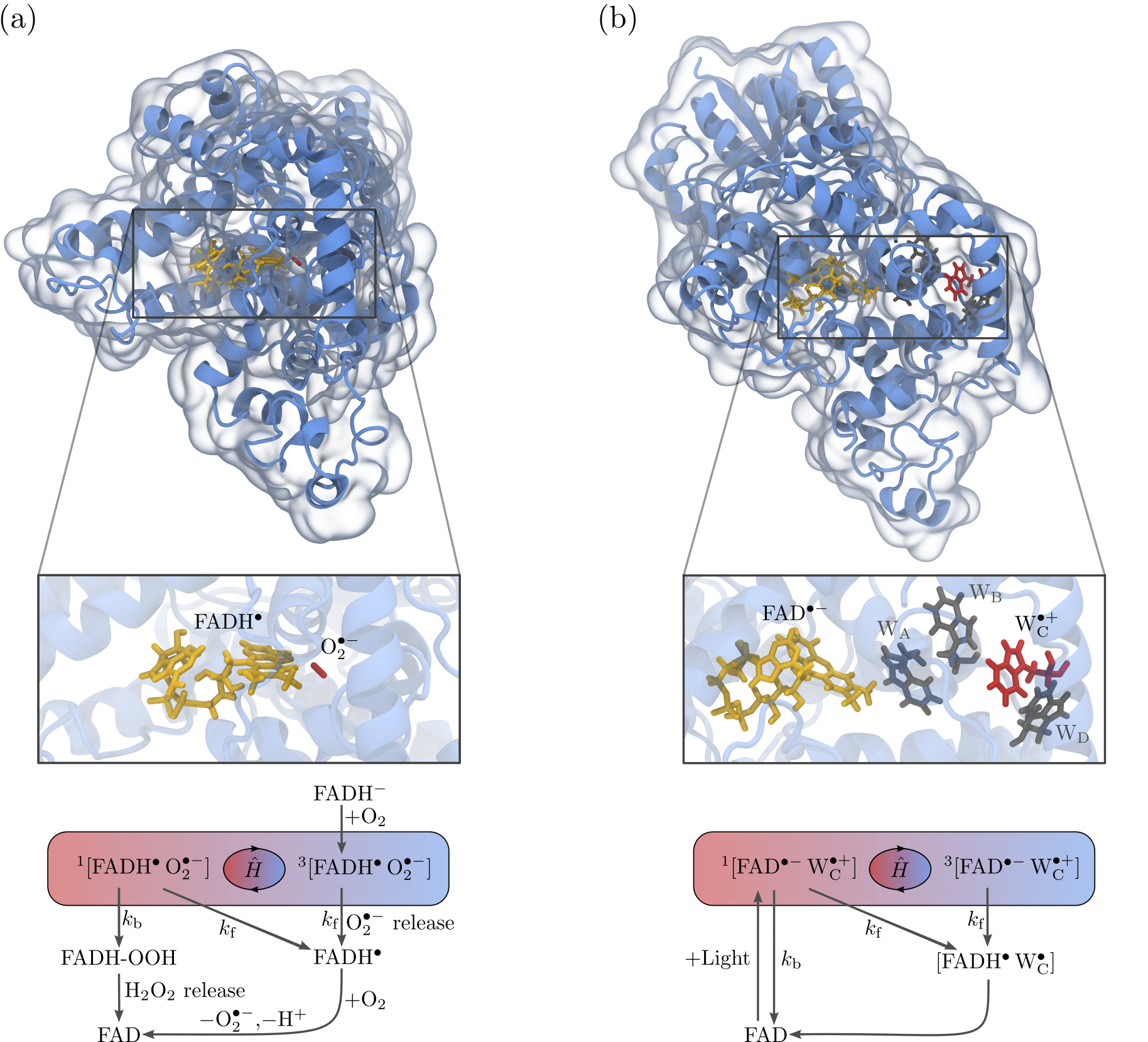}
     \caption{Avian (\textit{Columba livia}) cryptochrome 4a (ClCRY4a; PDB ID: 6pu0 \cite{zoltowskiChemicalStructuralAnalysis2019}) radical pair models and corresponding reaction schemes. a) shows the FADH$^{\bullet}$ and superoxide (O$^{\bullet-}_2$) radical pair and a corresponding reaction scheme, where the shaded box indicates magnetosensitive coherent interconversion between the singlet ($^{1}$[FADH$^{\bullet}$O$_{2}^{\bullet-}$]) and triplet ($^{3}$[FADH$^{\bullet}$O$_{2}^{\bullet-}$]) spin states under the Hamiltonian $\hat{H}$. b) shows the FAD$^{\bullet -}$ and W$_{\mathrm{C}}^{\bullet+}$ radical pair and a corresponding reaction scheme, where the shaded box indicates magnetosensitive coherent interconversion between the singlet ($^{1}$[FAD$^{\bullet -}$ W$_{\mathrm{C}}^{\bullet+}$]) and triplet ($^{3}$[FAD$^{\bullet -}$ W$_{\mathrm{C}}^{\bullet+}$]) spin states under the Hamiltonian $\hat{H}$. The reaction schemes apply to the canonical scenarios without CISS.}
     \label{fig:ciss_zeno_cry}
\end{figure*}

\section{Methods}

\subsection{Spin dynamics model}

We focus on the dependence of the recombination yield of the radical pair on the direction of the geomagnetic field. To this end, we consider the spin density operator $\hat{\rho}(t)$ of the radical pair system, which undergoes dynamics according to the master equation
\begin{align}
\dv{t} \hat{\rho}(t) &=  -\mathrm{i} \left[\hat{H}, \hat{\rho}(t)\right] - \left\{\hat{K}, \hat{\rho}(t)\right\} -k_{\mathrm{f}}\hat{\rho}(t) + \hat{\hat{\mathcal{R}}}_{\mathrm{NZ}} \hat{\rho}(t) \nonumber \\
&= \hat{\hat{\mathcal{L}}} \hat{\rho}(t) + \hat{\hat{\mathcal{R}}}_{\mathrm{NZ}} \hat{\rho}(t), 
\label{eq:NZ}
\end{align}
where $[ \cdot, \cdot ]$ and $\{ \cdot ,\cdot \}$ signify the commutator and an anti-commutator, respectively. In the second line, we have combined the action of coherent evolution according to the time-averaged spin Hamiltonian $\hat{H}$ ($\hbar = 1$), and reaction kinetics, accounted for by the recombination operator $\hat{K}$, into the Liouvillian term $\hat{\hat{\mathcal{L}}}$. For radical pair recombination in the singlet state, i.e.\ in the absence of CISS effects, $\hat{K}$ assumes the form
\begin{align}
\hat{K} = \frac{k_{\mathrm{b}}}{2} \hat{P}_{\mathrm{b}},
\label{eq:recombination_K}
\end{align}
where $k_{\mathrm{b}}$ is the recombination rate constant and $\hat{P}_{\mathrm{b}}$ is the recombination projection operator. In the presence of CISS, $\hat{P}_{\mathrm{b}}$ is equal to the singlet projection operator; its specific form in the presence of CISS is detailed below. In addition, we assumed that the signaling product is formed with a rate constant $k_{\mathrm{f}}$ independent of the spin configuration. The relaxation superoperator $\hat{\hat{\mathcal{R}}}_{\mathrm{NZ}}$, accounts for spin relaxation processes within the Nakajima-Zwanzig framework under the Markovian approximation, which has been shown to be more reliable than the Redfield theory when applied to asymmetrically recombining radical pairs \cite{fayElectronSpinRelaxation2019}. 

Specifically, for the coherent evolution governed by the Hamiltonian $\hat{H}$ we account for the Zeeman interaction with the external magnetic field, the hyperfine interaction between electrons and nuclei, and the electron-electron dipolar (EED) interradical coupling. These are defined as:
\begin{align}
    \hat{H}_{\rm Zee} =& -\gamma_e \sum_{i = 1,2} \hat{\vb{S}}_i \cdot \vb{B}, \\
    \hat{H}_{\rm HFC} =& \sum_{k=1}^{N_1} \hat{\textbf{I}}_{1,k} \cdot \textbf{A}_{1,k} \cdot \hat{\textbf{S}}_1 + \sum_{k=1}^{N_2} \hat{\textbf{I}}_{2,k} \cdot \textbf{A}_{2,k} \cdot \hat{\textbf{S}}_2, \\
    \hat{H}_{\rm EED} =& -d(r)(3(\hat{\vb{S}}_1 \cdot \vb{u})(\hat{\vb{S}}_2\cdot \vb{u}) - \hat{\vb{S}}_1 \cdot \hat{\vb{S}}_2),
    \label{eq:hamiltonian}
\end{align}
respectively, where $\hat{\vb{S}}$ and $\hat{\mathbf{I}}$ denote spin operators for electron and nuclear spins respectively, $\mathbf{A}$ is a hyperfine coupling tensor, and $\vb{u} = \vb{r}/r$ are elements of a unit vector along the dipolar axis with $r\equiv|\vb{r}|$. In our calculations, we transform the interradical axis to the CISS quantization frame (assumed to be the $z$ axis). $\gamma_e$ denotes the gyromagnetic ratio, and the dipolar coupling coefficient is $d(r) \equiv \mu_0g_e^2\mu_B^2/4\pi r^3 > 0$ with all constants having their usual meaning. The dipolar and hyperfine coupling tensors, obtained from refs.\,\cite{deviersAvianCryptochrome42024a, Gruning2022}, can be found in the Supplementary Information (SI). The precursor states before the effect of CISS for the FAD$^{\bullet -}$/W$_{\mathrm{C}}^{\bullet+}$ radical pair is the singlet state $\hat{\rho}_{\mathrm{S}}(0) = \hat{P}_{\mathrm{S}}/\mathrm{Tr}[\hat{P}_{\mathrm{S}}]$, where $\hat{P}_{\mathrm{S}}$ is the singlet projection operator, while the FADH$^{\bullet}$/O$^{\bullet-}_2$ radical pair is initialized in the triplet state $\hat{\rho}_{\mathrm{T}}(0) = \hat{P}_{\mathrm{T}}/\mathrm{Tr}[\hat{P}_{\mathrm{T}}]$, where $\hat{P}_{\mathrm{T}}$ is the triplet projection operator, as a consequence of molecular oxygen being a triplet molecule. In the precence of CISS, these initial states are modified, as detailed below. In Fig.~\ref{fig:ciss_zeno_cry} reaction schemes for each of these scenarios are shown.

To account for relaxation within the Nakajima-Zwanzig framework, we consider a system-bath interaction Hamiltonian of the form $\hat{H}_{I} = \sum_{i}X_{i}(t)\hat{A}_{i}$, where $\hat{A}_{i}$ is an operator in the system Hilbert space and $X_{i}$ is a random variable (satisfying $\langle X_{i} \rangle =0$) representing the protein's thermal motion and its coupling to the spin system. The relaxation superoperator can then be expressed as
\begin{align}
    \hat{\hat{\mathcal{R}}}_{\mathrm{NZ}} = -\sum_{j,k} \int_0^{\infty} g_{j,k}(\tau) \hat{\hat{\mathcal{A}}}_j^{\dag}  e^{\hat{\hat{\mathcal{L}}}\tau} \hat{\hat{\mathcal{A}}}_k \mathrm{d}\tau,
   \label{eq:Rnz}
\end{align}
where $g_{j,k}(t)$ is the time correlation function of $X_j$ and $X_k$ such that
\begin{equation}
g_{j,k}(t) = \left\langle X_j^*(0) X_k(t) \right\rangle
\end{equation}
and $\hat{\hat{\mathcal{A}}}_i\cdot = [\hat{A}_i,\cdot]$ denotes the commutation superoperator associated with $\hat{A}_i$.

As $\hat{\hat{\mathcal{L}}}$ can be expressed as $\hat{\hat{\mathcal{L}}} = -\mathrm{i} \left( \hat{H}_{\rm eff} \cdot - \cdot \hat{H}_{\rm{eff}}^{\dag}\right)$ with $\hat{H}_{\mathrm{eff}} = \hat{H} - \mathrm{i} \hat{K}$, it is possible to evaluate the equation of motion (Eq.~\ref{eq:NZ}) in the non-unitary eigenbasis of $\hat{H}_{\rm eff}$, for which Eq.~\ref{eq:Rnz} assumes the form of a sum over products of matrix elements of $\hat{A}_i$ and $\hat{A}_j$ multiplied by spectral densities
\begin{equation}
\mathcal{J}(\omega) = \int_0^\infty g(t) e^{\mathrm{i}\omega t} \mathrm{d}t,
\end{equation}
evaluated at the eigenvalue differences of $\hat{H}_{\rm eff}$. For random variables with exponential correlation functions, $g_{j,k}(t) = \left\langle X_j^* X_k \right\rangle \exp(-t/\tau_{\mathrm{c}})$, as assumed here for simplicity, $\mathcal{J}(\omega)$ assumes the form $\mathcal{J}_{j,k}(\omega) = \left\langle X_j^* X_k \right\rangle (\tau_{\mathrm{c}}^{-1} - \mathrm{i}\omega)^{-1}$. Here, $\tau_{\mathrm{c}}$ represents the correlation time, and $\left\langle X_j^* X_k \right\rangle$ is the covariance (mean squared fluctuation for $j=k$) of the system-bath couplings. The particular system-bath interaction we consider in this study is the random field relaxation (RFR), described by
\begin{align}
    \hat{H}_{\rm RFR}(t) = \sum_{i = x,y,z} \sum _{j = 1,2} b_{i,j}(t) \cdot \hat{S}_{i,j}, \label{eq:RFR}
\end{align}
which models stochastic fluctuations for $i \in \{x, y, z\}$ and for the two electrons in the radical pair ($j \in \{1,2\}$).

To assess the directional magnetic field effect of these radical pair systems, we evaluate the recombination yield, which, for a particular orientation ($\theta, \phi$) of the magnetic field, is given by
\begin{equation} 
    \Phi_\mathrm{b}(\theta,\phi) =  k_\mathrm{b}\int_0^\infty \mathrm{Tr}[\hat{P}_\mathrm{b}\,\hat{\rho}(t)]\mathrm{d}t.
\end{equation}
From this we vary the yield over $300$ independent magnetic field orientations to evaluate the maximal yield difference, referred to as the anisotropy
\begin{align}
    \Delta \Phi_{\mathrm{b}} = \Phi_{\mathrm{b}, \mathrm{MAX}} - \Phi_{\mathrm{b}, \mathrm{MIN}} 
\end{align}
and the sensitivity (relative anisotropy),
\begin{align}
    S = \frac{\Delta \Phi_{\mathrm{b}}}{\overline{\Phi}_{\mathrm{b}}},
\end{align}
where $\overline{\Phi}_{\mathrm{b}}$ denotes the mean of the recombination yield. To assess coherence, we calculate the relative entropy of coherence \cite{Baumgratz2014} with respect to the Zeeman basis, integrated over time, for both the ``local'' electron spin and ``global'' electron and nuclear spin system. For a directionally dependent measure we analyze both the mean coherence and maximal difference in coherence over magnetic-field orientations \cite{Smith2022}. Further details on the theoretical framework for assessing coherence are contained in the SI. Schematic diagrams of the models relating to FADH$^{\bullet}$/O$^{\bullet-}_2$ and FAD$^{\bullet -}$/W$_{\mathrm{C}}^{\bullet+}$ radical pairs are provided in Fig.~{\ref{fig:kb_kf_delta}}.

\subsection{CISS for singlet precursors}

We now turn our attention to the modeling of CISS. First, considering the case of the singlet-born radical pair FAD$^{\bullet -}$/W$_{\mathrm{C}}^{\bullet+}$, for the phenomenological chirality model of Luo and Hore \cite{Luo2021}, which we will refer to here as chirality-induced spin polarization (CISP), the state preparation and recombination are altered as follows. The chiral-mediated charge recombination reaction to a singlet product is described through eq.\ \eqref{eq:recombination_K} in combination with the CISP model recombination projection operator, indicated by a superscript P, given by
\begin{align}
    \hat{P}_{\chi}^{(\mathrm{P})} =& \vert \psi_{\chi}^{(\mathrm{P})}\rangle \langle \psi_{\chi}^{(\mathrm{P})} \vert \nonumber \\
    =& \frac{1}{4} - \hat{S}_{1z}\hat{S}_{2z} - (\hat{S}_{1x}\hat{S}_{2x} +  \hat{S}_{1y}\hat{S}_{2y})\cos(\chi)
    \nonumber \\&+ \frac{1}{2}(\hat{S}_{1z} - \hat{S}_{2z})\sin(\chi),
\end{align}
in which
\begin{align}
     \vert \psi_{\chi}^{(\mathrm{P})}\rangle = \cos(\frac{\chi}{2})\vert S \rangle + \sin(\frac{\chi}{2})\vert T_{0}\rangle, 
\end{align}
where $\vert S \rangle$ and $\vert T_{0} \rangle$ are the singlet and triplet state of zero spin projection, quantized with respect to the spin-polarization axis. The mixing angle $\chi$ parameterizes the extent of spin-polarization selectivity; maximal spin-polarization is realized for $\chi = \pi/2$ or $\chi= - \pi/2$, in which case the recombination requires the spin polarized states $\vert \uparrow \downarrow\rangle$ or $\vert \downarrow \uparrow \rangle$ of the donor and acceptor electrons, respectively. As the spin polarization is reversed for electron transfer traversing the chiral medium in the opposite direction, the associated charge separation reaction through the same chiral medium is described by $\hat{P}_{-\chi}^{(\mathrm{P})}$, i.e.\ the projector with negated $\chi$, favoring the opposite spin polarization. This further suggests that for a radical pair generated instantaneously from a singlet precursor in the chiral medium, the initial state is represented by   
\begin{align}
    \hat{\rho}_{\mathrm{S}}^{(\mathrm{P})}(0) =& \frac{1}{Z}\hat{P}_{-\chi}^{(\mathrm{P})} \nonumber \\ =& \frac{1}{Z}\bigg(\frac{1}{4} - \hat{S}_{1z}\hat{S}_{2z} - (\hat{S}_{1x}\hat{S}_{2x} +  \hat{S}_{1y}\hat{S}_{2y})\cos(\chi) \nonumber \\
    &- \frac{1}{2}(\hat{S}_{1z} - \hat{S}_{2z})\sin(\chi)\bigg). \label{eq:singlet_CISP}
\end{align}
A noteworthy consequence is that this model suggests a nonzero spin polarization $\langle \hat{S}_{1z} - \hat{S}_{2z} \rangle(0) = - \sin(\chi)$. 

Separately, in a first-principles derivation assuming an electron transfer reaction subject to non-vanishing spin-orbit transfer coupling, for singlet-born radical pairs, Fay has obtained a different reaction superoperator that predicts the emergence of chirality-induced spin coherence only \cite{fayChiralityInducedSpinCoherence2021}, which we here refer to as CISC. Employing the usual assumptions of Marcus theory, for a simple one-step charge recombination reaction, the reaction superoperator was found to involve a Haberkorn-like recombination term (on top of a superexchange-mediated spin-orbit coupling interaction) characterized by projector 
\begin{align}
    \hat{P}_{\theta}^{(\mathrm{C})} =& \vert \psi_{\theta}^{(\mathrm{C})} \rangle \langle \psi_{\theta}^{(\mathrm{C})}\vert \nonumber \\ =& \frac{1}{4} - \hat{S}_{1z}\hat{S}_{2z}
    - (\hat{S}_{1x}\hat{S}_{2x} + \hat{S}_{1y}\hat{S}_{2y})\cos(2\theta)\nonumber \\ &- (\hat{S}_{1x}\hat{S}_{2y} - \hat{S}_{1y}\hat{S}_{2x})\sin(2\theta),  
\end{align}
where the CISC model is indicated by superscript C and 
\begin{align}
    \vert \psi_{\theta}^{(\mathrm{C})} \rangle = \cos(\theta)\vert S \rangle + \mathrm{i}\sin(\theta)\vert T_{0} \rangle, 
\end{align}
with $\tan(\theta) = \Lambda/(2V)$ depending on the charge transfer coupling $V$ and the spin orbit coupling coefficient $\Lambda$. As before, the $z$-quantization axis coincides with the axis of the spin-orbit coupling, and a single electron transfer is considered. If the radical pair is formed very rapidly and irreversibly from a singlet precursor, the radical pair spin density operator is formed in the state
\begin{align}
    \hat{\rho}_{\mathrm{S}}^{(\mathrm{C})}(0) = \frac{1}{Z}\hat{P}_{\theta}^{\mathrm{(C)}}.
\end{align}
Unlike the CISP model, here the chiral electron transfer does not select spin polarization, i.e.\ $\langle \hat{S}_{1z}\rangle (0) = \langle \hat{S}_{2z}\rangle (0) = \langle \hat{S}_{1z} - \hat{S}_{2z}\rangle (0) = 0 $, but the imaginary part of the coherence between $\vert S \rangle$ and $\vert T_{0} \rangle $. Spin \emph{polarization} selectivity can only emerge in this picture if the model is extended to comprise a second electron transfer step, i.e.\ for a hopping mechanism, engaging an additional downstream state (e.g.\ bridge state) of non-zero exchange coupling and sufficient lifetime. Subsequently, the coherent evolution under the exchange interaction can convert the $\vert S \rangle$--$\vert T_{0}\rangle$ coherence of $\hat{\rho}_{\mathrm{S}}^{(\mathrm{C})}(0)$ into spin polarization in a process akin to the generation of electron spin polarization for chemically induced dynamic electron polarization (CIDEP) in diffusing radical pairs (i.e., $\vert S \rangle$--$\vert T_{0}\rangle$ interconversion in the separated pair interleaved by strong exchange coupling), as described by Kaptein \cite{kapteinChemicallyInducedDynamic1972}.  

\subsection{Interpolation and extension of CISS models}

For consideration of chiral-induced phenomena in the FADH$^{\bullet}$/O$^{\bullet-}_2$ radical pair, we need to account for triplet precursors. As these have not been considered in \cite{Luo2021} and \cite{fayOriginChiralityInduced2021}, we suggest a generalization.

Due to the phenomenological nature of the CISP model, it is \textit{a priori} difficult to predict a rigorous generalization to a triplet precursor. To attempt a resolution of this, we take inspiration from the postulated behavior of singlet precursors and the first-principles-derived CISC model. We note that the initial state $\hat{\rho}_{\mathrm{S}}^{\mathrm{(P)}}(0)$ cannot be obtained from the singlet precursor state $\hat{\rho}_{\mathrm{S}}(0) = \frac{1}{Z}\vert S \rangle \langle S \vert$, by a quantum channel that applies only to the transferred electron. However, one can obtain $\hat{\rho}_{\mathrm{S}}^{\mathrm{(P)}}(0)$ by combining a local phase rotation operation on the transferred electron in combination with evolution under exchange coupling, as suggested by Fay \cite{fayOriginChiralityInduced2021}, provided that the evolution time and exchange strength are chosen so that the family of states parametrized by the dephasing angle $\chi$ includes the maximally spin polarized state $\vert \uparrow \downarrow \rangle$. Specifically, denoting the Kraus operator for a phase rotation quantum channel by $\hat{W}=\vert \uparrow \rangle e^{\mathrm{i}\chi} \langle\uparrow \vert + \vert \downarrow \rangle \langle \downarrow \vert$, that acts on the transferred electron, and $\hat{U}_{\mathrm{ex}} = e^{\mathrm{i}j\hat{\mathbf{S}}_{1} \hat{\mathbf{S}}_{2}}$, the singlet state is transformed as
\begin{align}
    \hat{U}_{\mathrm{ex}}\hat{W}\hat{\rho}_{\mathrm{S}}(0)\hat{W}^{\dagger}\hat{U}_{\mathrm{ex}}^{\dagger} =& \frac{1}{Z}(\frac{1}{4} - \hat{S}_{1z}\hat{S}_{2z} \nonumber \\ 
    &- (\hat{S}_{1x}\hat{S}_{2x} + \hat{S}_{1y}\hat{S}_{2y})\cos(\chi) \nonumber \\
    &- (\hat{S}_{1x}\hat{S}_{2y} - \hat{S}_{1y}\hat{S}_{2x})\sin(\chi)\cos(4j) \nonumber \\
    &- \frac{1}{2}(\hat{S}_{1z} - \hat{S}_{2z})\sin(\chi)\sin(4j)).\label{eq:quantum_channel}
\end{align}
For $j=\frac{1}{4}(\frac{\pi}{2} + 2\pi n)$, with $n \in \mathbb{Z}$, this expression coincides with Luo and Hore's postulated singlet precursor state $\hat{\rho}_{\mathrm{S}}^{(\mathrm{P})}(0)$ (Eq.~\eqref{eq:singlet_CISP}) of the CISP model. 

With this established, we thus suggest that the triplet precursor for the CISP model can reasonably be assumed to be given by the quantum channel of Eq.~\eqref{eq:quantum_channel} acting instead on a (unpolarized) triplet initial state $\hat{\rho}_{\mathrm{T}}(0) = \hat{P}_{\mathrm{T}}/3Z$ and with $j=\frac{\pi}{8}$, thus yielding 
\begin{align}
    \hat{\rho}_{\mathrm{T}}^{(\mathrm{P})} =& \frac{1}{3Z}(\frac{1}{4} + \hat{S}_{1z}\hat{S}_{2z} + (\hat{S}_{1x}\hat{S}_{2x} + \hat{S}_{1y}\hat{S}_{2y})\cos(\chi) \nonumber \\ &+ \frac{1}{2}(\hat{S}_{1z} - \hat{S}_{2z})\sin(\chi)) \nonumber\\
    =& \frac{1}{3Z} - \frac{1}{3}\hat{\rho}_{\mathrm{S}}^{(\mathrm{P})}.
\end{align}

On the other hand, for the CISC model, an analogous derivation to the one given by Fay, but instead for a triplet precursor, yields
\begin{align}
        \hat{\rho}_{\mathrm{T}}^{(\mathrm{C})} =& \frac{1}{3Z}(\frac{1}{4} + \hat{S}_{1z}\hat{S}_{2z} + (\hat{S}_{1x}\hat{S}_{2x} + \hat{S}_{1y}\hat{S}_{2y})\cos(2\theta) \nonumber \\ &+ \frac{1}{2}(\hat{S}_{1x}\hat{S}_{2y} - \hat{S}_{1y}\hat{S}_{2x})\sin(2\theta)).
\end{align}
Note that $\hat{\rho}_{\mathrm{S}}^{(\mathrm{C})}(0)$ and $\hat{\rho}_{\mathrm{T}}^{(\mathrm{C})}(0)$ are also contained in the family of states described by the quantum channel in Eq.~\eqref{eq:quantum_channel}, for which they are recovered for $j=\frac{\pi}{2}n$, with $n \in \mathbb{Z}$ and $\chi=2\theta$, acting on the singlet and triplet precursors, respectively. Thus, Eq.~\eqref{eq:quantum_channel} can serve as an interpolant of the CISC and CISP approaches. In the results, we vary $\chi$ as the parameter that signifies the degree of CISS for both the CISP and the CISC models. In ref.~\cite{Poonia2023}, a different form of the chirality modulated triplet precursor is obtained based on a phenomenological spin selection process within the CISP framework, but not contained within the family of states described by Eq.~\eqref{eq:quantum_channel}, gives rise to a recombination operator involving a non-projector and thus lacks the corresponding connection to the CISC model.

\section{Results}

  \begin{figure*}
     \centering
     \includegraphics[width=\linewidth]{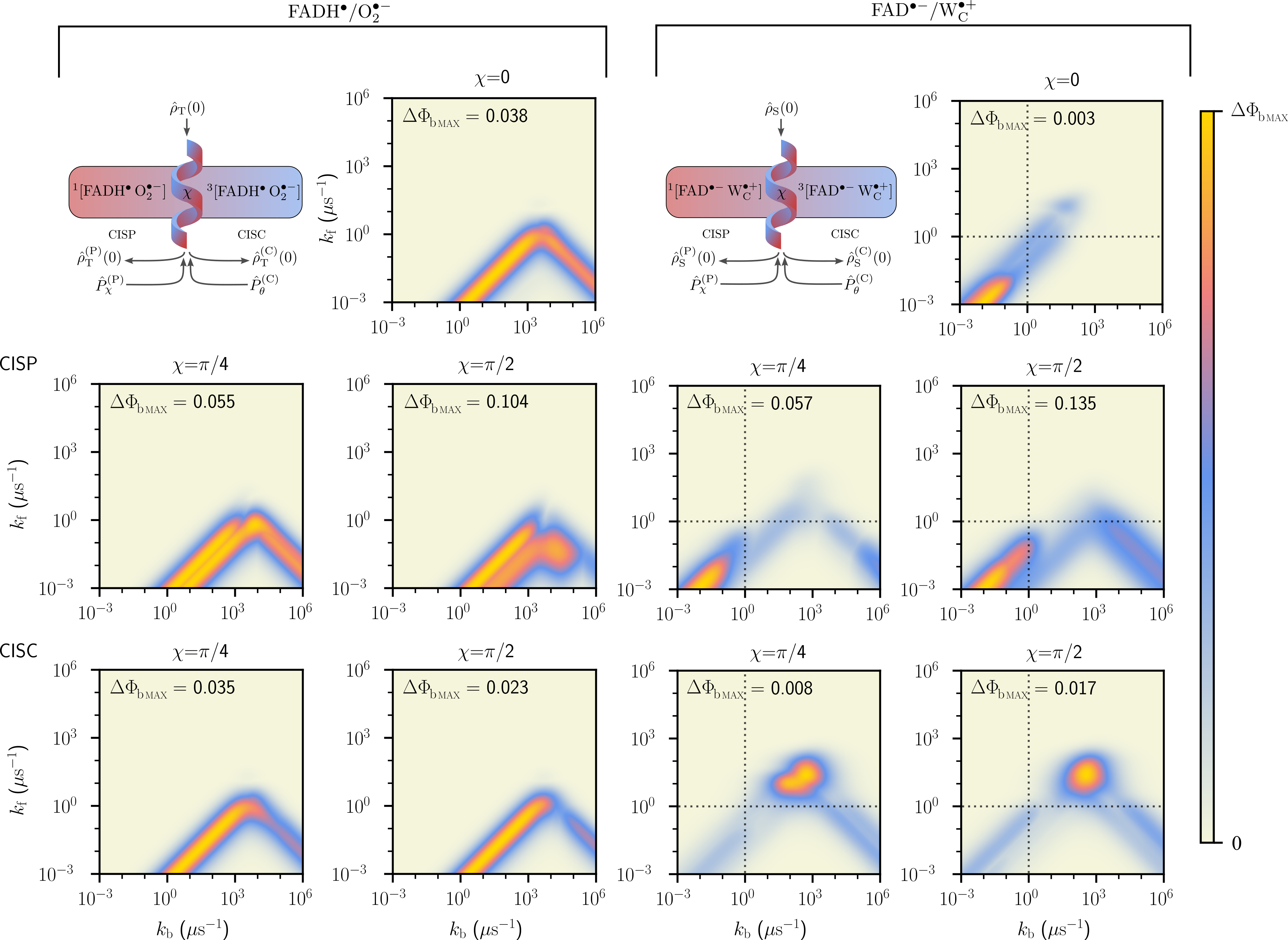}
     \caption{Heatmaps of anisotropy of the recombination yield $\Delta \Phi_{\mathrm{b}}$, sampled over $300$ magnetic field orientations and $200\times 200$ combinations of $k_{\mathrm{b}}$ and $k_{\mathrm{f}}$, for radical pairs subject to Zeeman, hyperfine, and EED interactions and CISS in the state preparation and recombination processes. Radical pair systems FADH$^{\bullet}$/O$^{\bullet-}_2$, comprising N$5$, and FAD$^{\bullet -}$/W$_{\mathrm{C}}^{\bullet+}$, comprising N$5$ of FAD and N$1$ of W$_{\mathrm{C}}$, are considered neglecting CISS effects ($\chi = 0$) and select chirality values $\pi/4,$ and $\pi/2$. As shown in the CISS reaction schemes, the initial states of these systems are modulated from the triplet precursor $\hat{\rho}_{\mathrm{T}}(0)$ to $\hat{\rho}_{\mathrm{T}}^{(\mathrm{P})}(0)$ and $\hat{\rho}_{\mathrm{T}}^{(\mathrm{C})}(0)$ for FADH$^{\bullet}$/O$^{\bullet-}_2$  and from the singlet precursor $\hat{\rho}_{\mathrm{S}}(0)$ to $\hat{\rho}_{\mathrm{S}}^{(\mathrm{P})}(0)$ and $\hat{\rho}_{\mathrm{S}}^{(\mathrm{C})}(0)$ for FAD$^{\bullet -}$/W$_{\mathrm{C}}^{\bullet+}$, where superscript P and C indicate the models of chiral induced spin polarization (CISP) and chiral induced spin coherence (CISC), respectively. The recombination projection operator $\hat{P}_{\mathrm{b}}$ is modulated to  $\hat{P}_{\theta}^{(\mathrm{C})}$, for $\chi=2\theta$ in the CISC model, and $\hat{P}_{\chi}^{(\mathrm{P})}$ for the CISP model. Dotted lines indicate common rate choices for the traditional FAD$^{\bullet -}$/W$_{\mathrm{C}}^{\bullet+}$ radical pair model. For ease of comparison, each heatmap has been normalized with the maximum value of $\Delta \Phi_{\mathrm{b}}$ labeled on each plot. }
     \label{fig:kb_kf_delta}
 \end{figure*}

  \begin{figure*}
     \centering
     \includegraphics[width=\linewidth]{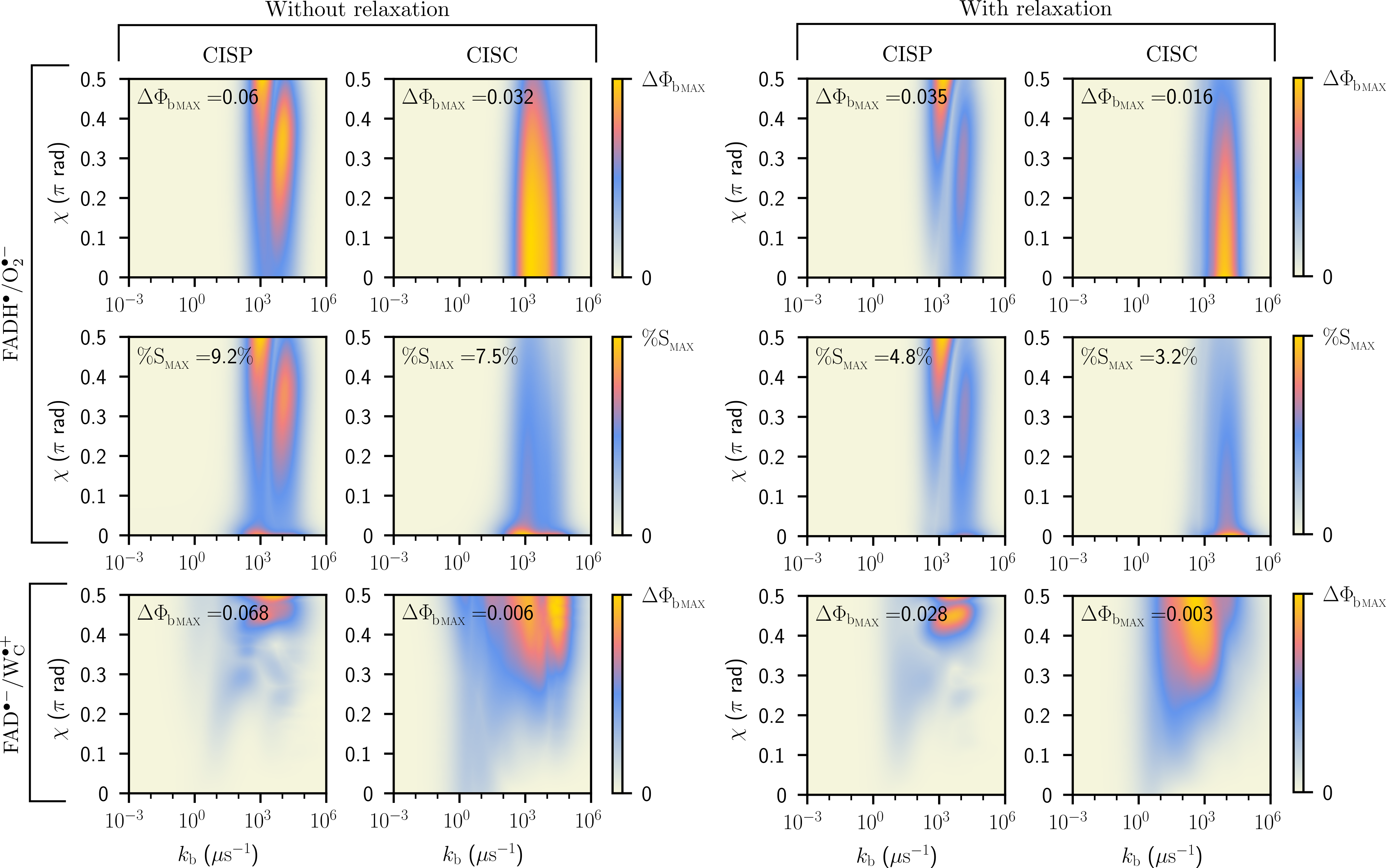}
     \caption{Heatmaps of anisotropy $\Delta \Phi_{\mathrm{b}}$, and percentage relative anisotropy \%$S$, sampled over $300$ magnetic field orientations of the recombination yield and $200\times 200$ combinations of $\chi$ and $k_{\mathrm{b}}$, with $k_{\mathrm{f}}=1\, \mu$s$^{-1}$. The radical pair systems and interactions considered are the same as in Fig.~\ref{fig:kb_kf_delta}. For comparison, each heatmap has been normalized with the maximum value of $\Delta \Phi_{\mathrm{b}}$ or \%$S$ indicated in each plot accordingly. On the left radical pair systems without spin relaxation are considered, whilst on the right relaxation due to uncorrelated random magnetic field fluctuations is included with an effective relaxation rate of $\gamma = 1\, \mu$s$^{-1}$.
     }
     \label{fig:chi_kb_sensitivities}
 \end{figure*}

 \begin{figure*}
     \centering
     \includegraphics[width=.9\linewidth]{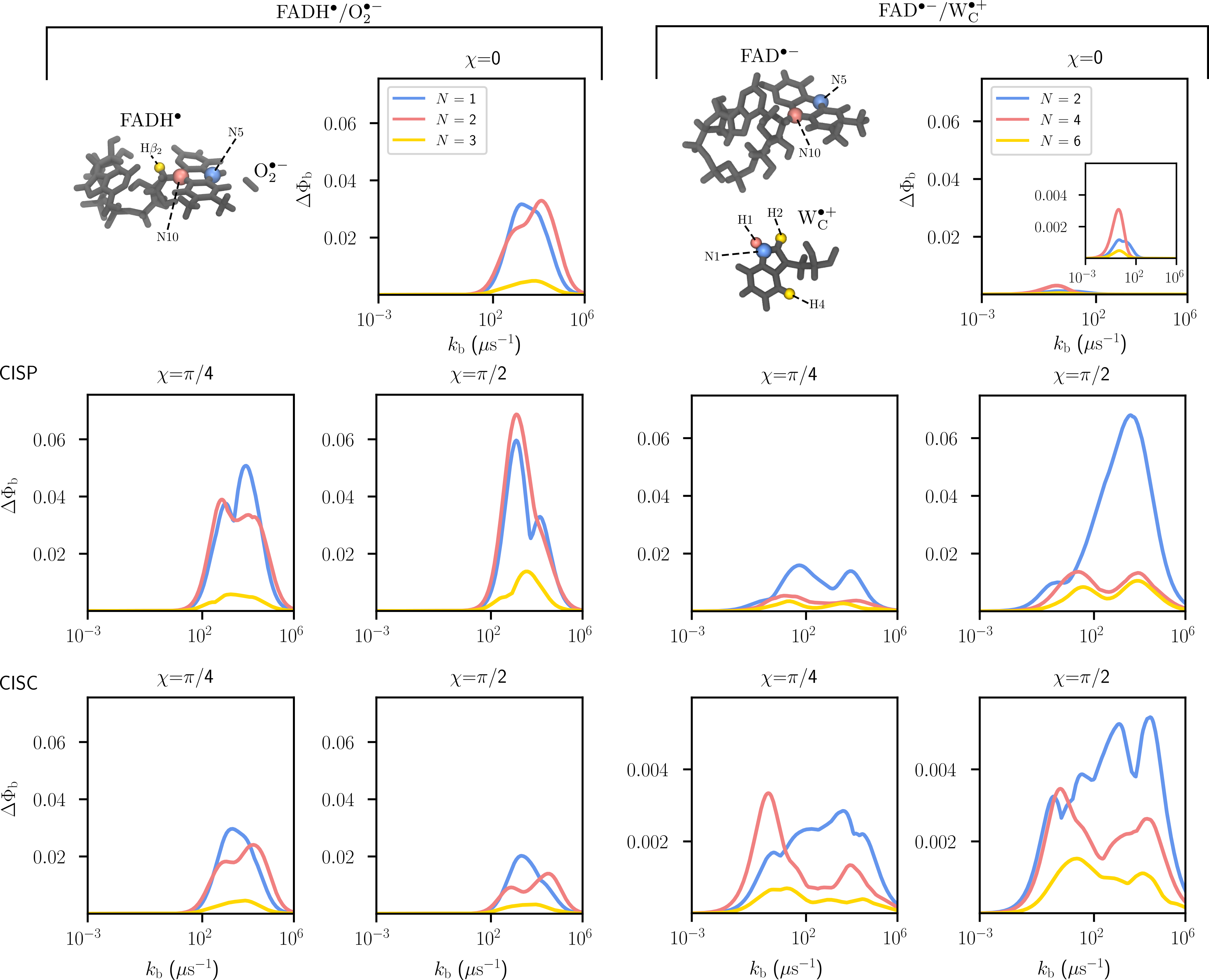}
     \caption{Line plots of anisotropy $\Delta \Phi_{\mathrm{b}}$, sampled over $300$ magnetic field orientations of the recombination yield for $200$ values of $k_{\mathrm{b}}$, with $k_{\mathrm{f}}=1\, \mu$s$^{-1}$, subject to Zeeman, hyperfine, and EED interactions. As in Figs.~\ref{fig:kb_kf_delta} and \ref{fig:chi_kb_sensitivities}, the case of neglecting CISS effects ($\chi = 0$) and  chirality in the state preparation and recombination processes are considered for $\chi=\pi/4,$ and $\chi=\pi/2$. To analyze persistence of effects, the number of hyperfine couplings is incremented to $N=3$ (in the order of N$5$, N$10$, and H$\beta2$) for FADH$^{\bullet}$/O$^{\bullet-}_2$ and in stages of $N=2$ (N$5$, N$1$), $N=4$ (incrementing by N$10$ and H$1$), and $N=6$ (incrementing by H$2$ and H$4$) for FAD$^{\bullet -}$/W$_{\mathrm{C}}^{\bullet+}$, as shown in the molecular diagrams. In the case of FAD$^{\bullet -}$/W$_{\mathrm{C}}^{\bullet+}$ the $y$--axis for the CISC model is matched to the inset to allow for qualitative comparison despite an order of magnitude drop in $\Delta \Phi_{\mathrm{b}}$.
  }
     \label{fig:persistence_delta}
 \end{figure*}
 
To understand the interrelation of the CISS and Zeno effects in enhancing the directional magnetic field effects in FAD$^{\bullet -}$/W$_{\mathrm{C}}^{\bullet+}$ and FADH$^{\bullet}$/O$^{\bullet-}_{2}$ in cryptochrome, we begin by assessing the anisotropy of the recombination yield, quantified by $\Delta \Phi_{\mathrm{b}}$, while varying the reaction rates for a total of $200\times 200$ combinations of $k_\mathrm{f}$ and $k_\mathrm{b}$ in the range of $10^{-3}$\,$\mu$s$^{-1}$ to $10^{6}$\,$\mu$s$^{-1}$. Fig.~\ref{fig:kb_kf_delta} shows the results for no CISS ($\chi=0$) and the two CISS models with $\chi=\pi/4$ and $\chi=\pi/2$, and $\chi = 2\theta$ to align CISC with CISP. Here, we consider minimally representative toy models of radical pairs comprising the dominant N$5$ hyperfine interaction of FAD and additionally the N$1$ interaction of W$_{\mathrm{C}}$ where it applies. First, for FADH$^{\bullet}$/O$^{\bullet-}_2$ in the absence of CISS ($\chi = 0$), we find the expected enhancement of the compass sensitivity brought about by highly asymmetric reaction rates with $k_\mathrm{b} \gg k_\mathrm{f}$, as attributed to the quantum Zeno effect and discussed in \cite{dentonMagnetosensitivityTightlyBound2024}. As $\chi$ increases for the CISP model, we observe an increase in maximal $\Delta \Phi_{\mathrm{b}}$ while the region of appreciable sensitivity is approximately conserved. For the CISC model, on the other hand, the sensitivity is generally attenuated, whereby the regions of larger $\Delta \Phi_{\mathrm{b}}$ are qualitatively similar as $\chi$ increases.

For FAD$^{\bullet -}$/W$_{\mathrm{C}}^{\bullet+}$, we observe no appreciable gain in $\Delta \Phi_{\mathrm{b}}$ across $k_\mathrm{f}$ and $k_\mathrm{b}$ with $\chi=0$. The model here favors balanced recombination rates with $k_\mathrm{f} \approx k_\mathrm{b}$. While the effects are generally small compared to FADH$^{\bullet}$/O$^{\bullet-}_{2}$, larger effects require long lifetimes. As $\chi$ increases in the CISP model, we observe an increase in $\Delta \Phi_{\mathrm{b}}$ emerging at asymmetric reaction rates, that is, showing the characteristic inverted V-profile of the quantum Zeno effect. In addition, a significant increase in compass sensitivity is present at symmetric rates ($k_\mathrm{f} \approx k_\mathrm{b}$), but this requires long coherent lifetimes of the radical pair ($k_\mathrm{f}$ and $k_\mathrm{b}$ $<0.1$\,$\mu$s$^{-1}$) and would therefore be exceedingly susceptible to attenuation/suppression by spin relaxation. For the CISC model, although enhancements emerge for rate combinations similar to those for the CISP model, an order of magnitude smaller gains are observed. Here, we have chosen to use $\Delta \Phi_{\mathrm{b}}$ for the point of comparison, as for FAD$^{\bullet -}$/W$_{\mathrm{C}}^{\bullet+}$ the relative anisotropy is dominated by small values of the mean yield $\overline{\Phi}_{\mathrm{b}}$. However, we have presented plots of \%$S$ for FADH$^{\bullet}$/O$^{\bullet-}_2$ in the SI (Fig.~S1).

Motivated by the findings on $k_\mathrm{f}$ and $k_\mathrm{b}$, we set out to evaluate the optimal CISS degree, as encoded by $\chi$, where in the CISC model $\chi = 2\theta$, for $k_\mathrm{f}=1$\,$\mu$s$^{-1}$. This forward rate constant has been chosen because it agrees with the order of magnitude of the experimental radical pair lifetime in \textit{in vitro} experiments \cite{maedaMagneticallySensitiveLightinduced2012} and the expected spin relaxation rates \cite{gruningEffectSpinRelaxation2024, Player2019, Kattnig2016a, karogodinaKineticMagneticfieldEffect2011, Hogben2009} thus focusing on a situation for which the anisotropy loss due to spin relaxation is likely tolerable, unlike for the much smaller $k_\mathrm{f}$ values considered in Fig.~\ref{fig:kb_kf_delta}. In contrast, we vary $k_\mathrm{b}$ as it is plausible that the recombination rate constant extends to a larger range of values. Specifically, in Fig.~\ref{fig:chi_kb_sensitivities} we considered $200\times 200$ combinations of $0\leq\chi\leq 0.5\pi$ and $10^{-3} \leq k_{\mathrm{b}} \leq 10^{6}$\,$\mu$s$^{-1}$, for CISP and CISC models of the radical pair systems. First, considering CISP for FADH$^{\bullet}$/O$^{\bullet-}_2$, we observe that an increase in $\chi$ can lead to an increase in $\Delta \Phi_{\mathrm{b}}$, interpolating between the two bands of $k_\mathrm{b}$ values for which the quantum Zeno effect emerges. Furthermore, by evaluating \%$S$ we observe that, while enhancements by increasing $\chi$ remain, it is also possible to achieve similar increases in \%$S$ via the quantum Zeno effect with $\chi=0$, provided that the recombination is strongly asymmetric ($k_{\mathrm{b}} \approx 10^{3}$\,$\mu$s$^{-1})$. Conversely, for the CISC model it is demonstrated that an increase in CISS-generated coherence only diminishes the enhancement provided by the quantum Zeno effect. 

In comparison, for the CISP model, FAD$^{\bullet -}$/W$_{\mathrm{C}}^{\bullet+}$ demonstrates a significant increase in $\Delta \Phi_{\mathrm{b}}$ as the CISS effect becomes large and approaches $\chi=\pi/2$. However, this once again occurs contingent on the presence of asymmetric reaction rates with $k_{\mathrm{b}}\approx 10^{3}$\,$\mu$s$^{-1})$. Similar observations emerge for the CISC model, but the increases in $\Delta \Phi_{\mathrm{b}}$ are minor, only achieving up to an order of magnitude less than that predicted in the CISP model. 

To assess the robustness of enhancements to noise, we have also considered the inclusion of spin relaxation using the RFR noise model incorporated via the Nakajima-Zwanzig formalism (Eqs.~\eqref{eq:NZ}, \eqref{eq:Rnz}, and \eqref{eq:RFR}), in which the two electron spins experience uncorrelated magnetic field noise of equal amplitude in the three Cartesian directions. Although RFR attenuates magnetosensitivity to approximately half its magnitude, the enhancements persist in both the CISP and CISC models at $k_{\mathrm{b}}$ and $\chi$ values similar to those observed without relaxation. This is shown in Fig.~\ref{fig:chi_kb_sensitivities} for an effective relaxation rate $\gamma = \langle \delta B_{i}^{2} \rangle \tau_{\mathrm{c}} = 1\, \mu$s$^{-1}$, where $\langle \delta B_{i}^{2} \rangle$ denotes the variance of the Larmor precession frequency due to the random field, and the correlation time is taken as $\tau_{\mathrm{c}}=1\,$ns. Within the CISP model the most robust $\Delta \Phi_{\mathrm{b}}$ and \%$S$ values are observed as $\chi$ is increased, and similarly for the CISP and CISC models with the FAD$^{\bullet -}$/W$_{\mathrm{C}}^{\bullet+}$ radical pair. In contrast, for the CISC model with the FADH$^{\bullet}$/O$^{\bullet-}_2$ radical pair, the most robust sensitivities are observed for small values of $\chi$. Overall, enhancements remain restricted to regimes of strongly asymmetric recombination.

Finally, we increase the complexity of the radical pair model by incrementing the number of hyperfine interactions, in order of the most dominant anisotropic interactions, to assess whether CISS and quantum Zeno enhancements persist in more realistic models. Specifically, in Fig.~\ref{fig:persistence_delta}, we once again set $k_\mathrm{f}=1$\,$\mu$s$^{-1}$ and evaluate 
$\Delta \Phi_{\mathrm{b}}$ in the CISP and CISC models for $200\times 200$ combinations of $0\leq\chi\leq 0.5\pi$ and $10^{-3} \leq k_{\mathrm{b}} \leq 10^{6}\,\mu$s$^{-1}$. For FADH$^{\bullet}$/O$^{\bullet-}_2$, we increase the number of FAD hyperfine coupled nuclei to $N=3$ (in the order of N$5$, N$10$, and H$\beta2$). For FAD$^{\bullet -}$/W$_{\mathrm{C}}^{\bullet+}$, interactions in both radicals are relevant, and so we use stages of $N=2$ (N$5$, N$1$), $N=4$ (incrementing by N$10$ and H$1$),  with interactions successively added in both radicals, and lastly $N=6$ (incrementing by H$2$ and H$4$ of W$_\mathrm{C}^{\bullet +}$). Parameter values for the hyperfine couplings used are provided in the SI. For FADH$^{\bullet}$/O$^{\bullet-}_2$ with $\chi=0$ we observe that the quantum Zeno enhancement of $\Delta \Phi_{\mathrm{b}}$ demonstrates reasonable resilience as the number of hyperfine couplings increases, as was also shown previously in ref.~\cite{dentonMagnetosensitivityTightlyBound2024}, for asymmetric reaction rates in the approximate range $10^{2}\leq k_{\mathrm{b}}\leq 10^{6}\,\mu$s$^{-1}$. In comparison, the CISP model predicts a further increase in $\Delta \Phi_{\mathrm{b}}$ as $\chi$ increases, at similar asymmetric rate choices. In addition to this base enhancement, present even at $N=1$, no further resilience is observed to an increase in hyperfine-coupled nuclei in the presence of CISS. In contrast, the CISC model predicts a decrease in quantum Zeno enhancement as $\chi$ increases. 

For FAD$^{\bullet -}$/W$_{\mathrm{C}}^{\bullet+}$ with $\chi=0$, the highest $\Delta \Phi_{\mathrm{b}}$ is predicted for rates closer to symmetric, but is still small and continues to drop as the number of hyperfine couplings increases. However, for the CISP model, as $\chi$ increases, an enhancement emerges, becoming significant for asymmetric rates in the range $1<k_{\mathrm{b}}\leq 10^{6}\, \mu$s$^{-1}$. This effect appears to exhibit a resilience to increasing the number of hyperfine couplings comparable to that observed for the quantum Zeno-enhanced sensitivities in FADH$^{\bullet}$/O$^{\bullet-}_2$. Although qualitatively similar observations emerge for the CISC model, enhancements are minor and an order of magnitude smaller than predicted by the CISP model.

\section{Discussion}
  \begin{figure*}
     \centering
     \includegraphics[width=\linewidth]{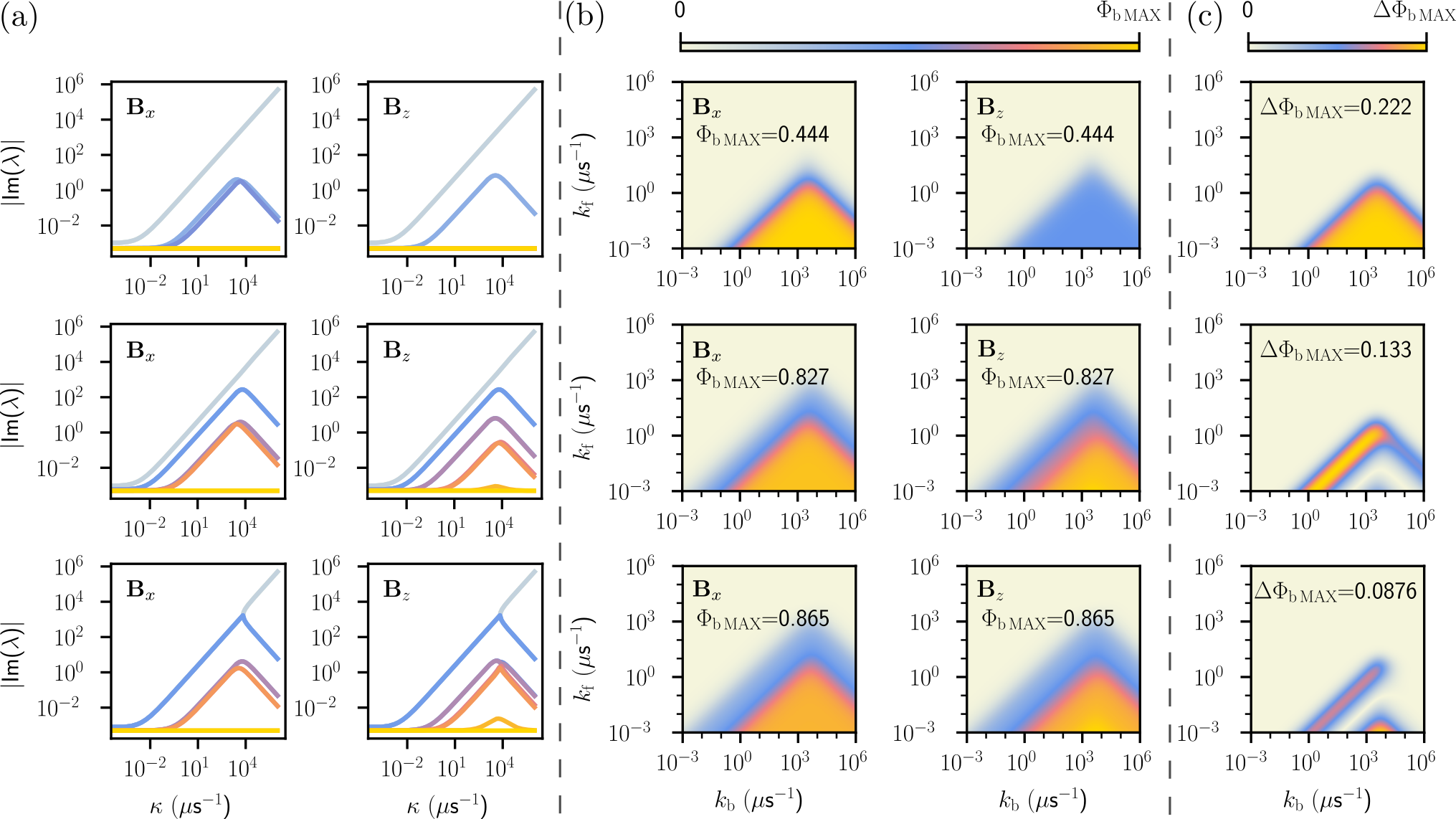}
     \caption{Comparison of eigenvalues of the effective system Hamiltonian, to recombination yields and anisotropy for a toy FADH$^{\bullet}$/O$^{\bullet-}_{2}$ radical pair and a single hyperfine coupling with $A_{xx}=A_{yy}=A_{\perp}=0$ and $A_{zz}=A_{\parallel}>>\omega$, subject to Zeeman, hyperfine, and EED interactions. CISS effects are absent with $\chi=0$ (first row), and included within the CISP model for $\chi=\pi/4$ (second row) and $\chi=\pi/2$ (third row). (a) plots of of the absolute value of the $12$ eigenvalues of the effective system Hamiltonian, (b) heatmaps of recombination yield $\Phi_{\mathrm{b}}$ for $200\times 200$ combinations of $k_{\mathrm{b}}$ and $k_{\mathrm{f}}$, and (c) heatmaps of anisotropy $\Delta\Phi_{\mathrm{b}}$ are shown for contrasting applied magnetic field directions $\mathrm{\mathbf{B}}_{x}$ and $\mathrm{\mathbf{B}}_{z}$, as defined in the molecular frame. 
  }
     \label{fig:evalue_toy}
 \end{figure*}
Our understanding of CISS generation in dyad molecules is still in its infancy. In the context of radical pairs in cryptochrome-based magnetoreception, only the phenomenological CISP model introduced by Luo and Hore has been previously studied. Although these studies capture core expectations of a CISS effect, the model itself is not based upon microscopic derivation, nor does it have supporting experimental evidence. Alternative models, such as spin filtering-based mechanisms \cite{chiesaAssessingNatureChiralInduced2021}, are conceivable. Here, we have demonstrated that an alternative CISC model, introduced by Fay based on a microscopic derivation for a one-step electron transfer and inducing spin coherence only, is insufficient to significantly increase the sensitivity of prototypical radical pairs to the direction of the geomagnetic field, and can even be detrimental, suggesting that polarization is an essential feature of the enhancements observed previously. We note that the CISC model does not coincide with the CISP model, even if a polarization-generating hopping process is assumed. However, with hopping and exchange, the models share common features, such that in both models the initial state comprises coherent singlet-triplet mixing and the emergence of spin polarization. Moreover, we have found that both approaches emerge from a quantum channel description of these processes for different limiting cases of the exchange interaction. In the context of donor-bridge-acceptor triads, Fay and Limmer have derived a comprehensive master equation for CISS-modulated recombination, which is parameterized by the various kinetic rate parameters of the triad \cite{faySpinSelectiveCharge2023}. It is therefore possible that extensions to the CISC model \cite{fayOriginChiralityInduced2021}, which have demonstrated validity in interpreting experimental results in photosystem I \cite{carmeliSpinSelectivityElectron2014} and donor-acceptor molecules \cite{eckvahlDirectObservationChiralityinduced2023}, can replicate the conditions for enhancements to emerge.

A prominent gap remains even within reported observations of CISS enhancements to sensitivity and coherence in cryptochrome insofar as they lack an explanation for their origin. By identifying that enhancements occur for asymmetric reaction kinetics, and comparing the singlet-born FAD$^{\bullet -}$/W$_{\mathrm{C}}^{\bullet+}$ radical pair, and extending the theory to accommodate for the triplet-born FADH$^{\bullet}$/O$^{\bullet-}_2$ radical pair, our results indicate that enhancements are to some extent, arguably predominantly, due to the quantum Zeno effect. To establish this further, we have considered a FADH$^{\bullet}$/O$^{\bullet-}_2$ toy radical pair model, subject to Zeeman, hyperfine, and EED interactions, in which there is a single hyperfine coupling satisfying $A_{xx}=A_{yy}=A_{\perp}=0$ and $A_{zz}=A_{\parallel}>>\omega$, where $\omega$ is the Larmor precession frequency in the geomagnetic field. In Fig.~\ref{fig:evalue_toy} the eigenvalues of the effective Hamiltonian are compared to the recombination yields and anisotropy, for contrasting applied magnetic field directions of $\mathbf{B}_{x}$ and $\mathbf{B}_{z}$, and the cases of neglecting CISS ($\chi=0$), and including CISS effects under the CISP model with $\chi=\pi/4$ and $\chi = \pi/2$. These plots show that the underlying magnetosensitivity enhancement originates from a contrast in the recombination yield enabled by eigenstate couplings that are engaged either when the field is along $x$ or $z$ (here defined with respect to the molecular frame) and exhibit a $1/k_{\mathrm{b}}$ scaling in the imaginary parts of the eigenvalues. This is most clearly shown in the close correspondence between the profile of $\Delta \Phi_{\mathrm{b}}$ and eigenstate coupling splitting. In ref.~\cite{dentonMagnetosensitivityTightlyBound2024} it was demonstrated that these features are signatures of the quantum Zeno effect. 

Regarding the persistence of sensitivity enhancements, as the number of hyperfine coupled nuclei per radical is increased, the quantum Zeno-enhanced sensitivity persists, though subject to an expected overall drop in magnitude. The sensitivity is comparable in both radical pairs, whether that is by CISS-reinforced Zeno effects in FADH$^{\bullet}$/O$^{\bullet-}_2$, or CISS-activated Zeno effects in FAD$^{\bullet -}$/W$_{\mathrm{C}}^{\bullet+}$ by adding an initial triplet state characteristic (cf.\ Eq.~\eqref{eq:singlet_CISP}) \cite{katsoprinakisCoherentTripletExcitation2010}. Whilst large enhancements for the FAD$^{\bullet -}$/W$_{\mathrm{C}}^{\bullet+}$ radical pair can be achieved at low reaction rate regimes, the long radical pair lifetimes required for this effect likely negate its realization due to the presence of significant spin relaxation. However, the magnetosensitivity was demonstrated to be reasonably robust to random field relaxation with an effective relaxation rate of $\gamma = 1\,\mu$s$^{-1}$ predominantly at strongly asymmetric recombination regimes where the quantum Zeno effect is operative.

The observations here qualitatively carry over to coherence enhancements, with regard to significant effects predominantly emerging for largely asymmetric rates, and thus we have reserved these results for the SI (Figs.~S2--S11). However, there are some key observations that are worth discussing. First, there is a close correspondence of $\Delta \Phi_{\mathrm{b}}$ and the maximal difference in coherence over magnetic field orientations; this is present even for $\chi=0$ and is also a feature of the quantum Zeno effect as demonstrated in the SI of ref.~\cite{dentonMagnetosensitivityTightlyBound2024}. The mean value of coherence does not necessarily accompany $\Delta \Phi_{\mathrm{b}}$, signifying that variability in coherence is a better indicator of sensitivity than large average coherence. With increase in the number of hyperfine couplings, the mean coherence demonstrates remarkable resilience, though this is also true for the quantum Zeno regime without chirality in FADH$^{\bullet}$/O$^{\bullet-}_2$. Similar profiles emerge for both the CISP and CISC models, further indicating that coherence enhancement does not always imply sensitivity enhancement. It is important to note that we have chosen to calculate coherence in the Zeeman basis to allow comparison with previous observations of coherence enhancements related to CISS \cite{Poonia2023}. 

Returning to the principal questions of this investigation, we find that CISS can conditionally lead to enhanced magnetosensitivity and coherence in cryptochrome, that it does so predominantly by influencing features that, based on their characteristic dependence on asymmetric reaction rates, ought to be attributed to the quantum Zeno effect, and that it is effective only if polarization is generated. If only coherence is generated, then significant enhancements do not emerge and CISS can even be detrimental by diminishing enhancements gained through the quantum Zeno effect, though not fully negating them. Previously we found that the quantum Zeno effect can reinstate sensitivity for a tightly bound triplet-born FADH$^{\bullet}$/O$^{\bullet-}_2$ radical pair \cite{dentonMagnetosensitivityTightlyBound2024}, thought to be infeasible due to strong interradical interactions or spin relaxation. Here, we find that CISS generated polarization can amplify this effect for radical pairs with triplet precursors and extend it to singlet precursors. The role of CISS in observed enhancements is thus revealed as modifying the required initial-state triplet characteristics and recombination for the quantum Zeno effect to be bolstered or enabled. Based on our models, experimental confirmation of CISS generated polarization in cryptochrome-based radical pairs could suggest that it acts as a functional resource, actively utilized and potentially evolutionarily adapted by nature for magnetoreception, whereas its absence could categorize CISS as a spandrel (when indifferent) or even a biological constraint (in scenarios in which it is detrimental). Beyond this, spin-polarization mechanisms have been proposed more broadly in the context of photo-induced electron transfer reactions \cite{fayOriginChiralityInduced2021}, provided there is an intermediate state, suggesting a potential wider-scale emergence of the quantum Zeno effect beyond magnetoreception. In particular, its role in donor-chiral-bridge-acceptor systems \cite{linMolecularEngineeringEmissive2025, maniMolecularQubitsBased2022} should be further investigated to determine the potential for a chirality-mediated quantum Zeno effect not only in biological processes, but also for manipulation alongside quantum control \cite{chowdhuryQuantumControlRadicalPair2024} in emerging molecular quantum technologies. 


\section*{Supplementary Material}
See the supplementary material for additional simulation details, including spin Hamiltonian parameters, and a discussion of coherence measures.

\begin{acknowledgments}
The authors thank the Engineering and Physical Sciences Research Council (EPSRC grants EP/V047175/1 and EP/X027376/1), the Biotechnology and Biological Sciences Research Council (BBSRC grants BB/Y514147/1, BB/Y51312X/1), and the Office of Naval Research Global (ONR-G Award Number N62909-21-1-2018) for support. The authors further acknowledge the use of the University of Exeter High Performance Computing Facility. For the purpose of open access, the authors have applied a Creative Commons Attribution (CC BY) license to any Author Accepted Manuscript version arising from this submission.
\end{acknowledgments}

\section*{Author Declarations}
\subsection*{Conflict of interest}
The authors have no conflicts to disclose.
\subsection*{Author contributions}
D.R.K.\ conceived the study. All authors contributed to the research through conducting computer simulation, data analysis, or analytic derivations. L.D.S., S.T., M.C.J.D.\ and D.R.K. developed the underpinning software. D.R.K.\ supervised the work. L.D.S.\ wrote the manuscript, with inputs from co-authors. D.R.K.\ acquired the financial support for the project, and managed and coordinated the research activities.

\section*{Data Availability Statement}
The data that support the findings of this study are available within the article and its supplementary material.

\bibliography{chirality_zeno_tidy.bib}

\begin{thebibliography}{82}%
\makeatletter
\providecommand \@ifxundefined [1]{%
 \@ifx{#1\undefined}
}%
\providecommand \@ifnum [1]{%
 \ifnum #1\expandafter \@firstoftwo
 \else \expandafter \@secondoftwo
 \fi
}%
\providecommand \@ifx [1]{%
 \ifx #1\expandafter \@firstoftwo
 \else \expandafter \@secondoftwo
 \fi
}%
\providecommand \natexlab [1]{#1}%
\providecommand \enquote  [1]{``#1''}%
\providecommand \bibnamefont  [1]{#1}%
\providecommand \bibfnamefont [1]{#1}%
\providecommand \citenamefont [1]{#1}%
\providecommand \href@noop [0]{\@secondoftwo}%
\providecommand \href [0]{\begingroup \@sanitize@url \@href}%
\providecommand \@href[1]{\@@startlink{#1}\@@href}%
\providecommand \@@href[1]{\endgroup#1\@@endlink}%
\providecommand \@sanitize@url [0]{\catcode `\\12\catcode `\$12\catcode `\&12\catcode `\#12\catcode `\^12\catcode `\_12\catcode `\%12\relax}%
\providecommand \@@startlink[1]{}%
\providecommand \@@endlink[0]{}%
\providecommand \url  [0]{\begingroup\@sanitize@url \@url }%
\providecommand \@url [1]{\endgroup\@href {#1}{\urlprefix }}%
\providecommand \urlprefix  [0]{URL }%
\providecommand \Eprint [0]{\href }%
\providecommand \doibase [0]{https://doi.org/}%
\providecommand \selectlanguage [0]{\@gobble}%
\providecommand \bibinfo  [0]{\@secondoftwo}%
\providecommand \bibfield  [0]{\@secondoftwo}%
\providecommand \translation [1]{[#1]}%
\providecommand \BibitemOpen [0]{}%
\providecommand \bibitemStop [0]{}%
\providecommand \bibitemNoStop [0]{.\EOS\space}%
\providecommand \EOS [0]{\spacefactor3000\relax}%
\providecommand \BibitemShut  [1]{\csname bibitem#1\endcsname}%
\let\auto@bib@innerbib\@empty
\bibitem [{\citenamefont {Wiltschko}\ and\ \citenamefont {Wiltschko}(2019)}]{Wiltschko2019}%
  \BibitemOpen
  \bibfield  {author} {\bibinfo {author} {\bibfnamefont {R.}~\bibnamefont {Wiltschko}}\ and\ \bibinfo {author} {\bibfnamefont {W.}~\bibnamefont {Wiltschko}},\ }\bibfield  {title} {\bibinfo {title} {Magnetoreception in birds},\ }\href {https://doi.org/10.1098/RSIF.2019.0295} {\bibfield  {journal} {\bibinfo  {journal} {J. R. Soc. Interface}\ }\textbf {\bibinfo {volume} {16}},\ \bibinfo {pages} {20190295} (\bibinfo {year} {2019})}\BibitemShut {NoStop}%
\bibitem [{\citenamefont {Mouritsen}(2018)}]{Mouritsen2018}%
  \BibitemOpen
  \bibfield  {author} {\bibinfo {author} {\bibfnamefont {H.}~\bibnamefont {Mouritsen}},\ }\bibfield  {title} {\bibinfo {title} {Long-distance navigation and magnetoreception in migratory animals},\ }\href {https://doi.org/10.1038/s41586-018-0176-1} {\bibfield  {journal} {\bibinfo  {journal} {Nature}\ }\textbf {\bibinfo {volume} {558}},\ \bibinfo {pages} {50} (\bibinfo {year} {2018})}\BibitemShut {NoStop}%
\bibitem [{\citenamefont {Johnsen}\ and\ \citenamefont {Lohmann}(2008)}]{Johnsen2008}%
  \BibitemOpen
  \bibfield  {author} {\bibinfo {author} {\bibfnamefont {S.}~\bibnamefont {Johnsen}}\ and\ \bibinfo {author} {\bibfnamefont {K.~J.}\ \bibnamefont {Lohmann}},\ }\bibfield  {title} {\bibinfo {title} {Magnetoreception in animals},\ }\href {https://doi.org/10.1063/1.2897947} {\bibfield  {journal} {\bibinfo  {journal} {Phys. Today}\ }\textbf {\bibinfo {volume} {61}},\ \bibinfo {pages} {29} (\bibinfo {year} {2008})}\BibitemShut {NoStop}%
\bibitem [{\citenamefont {Hore}\ and\ \citenamefont {Mouritsen}(2016)}]{Hore2016}%
  \BibitemOpen
  \bibfield  {author} {\bibinfo {author} {\bibfnamefont {P.~J.}\ \bibnamefont {Hore}}\ and\ \bibinfo {author} {\bibfnamefont {H.}~\bibnamefont {Mouritsen}},\ }\bibfield  {title} {\bibinfo {title} {The radical-pair mechanism of magnetoreception},\ }\href {https://doi.org/10.1146/ANNUREV-BIOPHYS-032116-094545} {\bibfield  {journal} {\bibinfo  {journal} {Annu. Rev. Biophys.}\ }\textbf {\bibinfo {volume} {45}},\ \bibinfo {pages} {299} (\bibinfo {year} {2016})}\BibitemShut {NoStop}%
\bibitem [{\citenamefont {Alvarez}\ \emph {et~al.}(2024)\citenamefont {Alvarez}, \citenamefont {Gerhards}, \citenamefont {Solov’yov},\ and\ \citenamefont {de~Oliveira}}]{alvarezQuantumPhenomenaBiological2024}%
  \BibitemOpen
  \bibfield  {author} {\bibinfo {author} {\bibfnamefont {P.~H.}\ \bibnamefont {Alvarez}}, \bibinfo {author} {\bibfnamefont {L.}~\bibnamefont {Gerhards}}, \bibinfo {author} {\bibfnamefont {I.~A.}\ \bibnamefont {Solov’yov}},\ and\ \bibinfo {author} {\bibfnamefont {M.~C.}\ \bibnamefont {de~Oliveira}},\ }\bibfield  {title} {\bibinfo {title} {Quantum phenomena in biological systems},\ }\bibfield  {journal} {\bibinfo  {journal} {Front. Quantum Sci. Technol.}\ }\textbf {\bibinfo {volume} {3}},\ \href {https://doi.org/10.3389/frqst.2024.1466906} {10.3389/frqst.2024.1466906} (\bibinfo {year} {2024})\BibitemShut {NoStop}%
\bibitem [{\citenamefont {Kim}\ \emph {et~al.}(2021)\citenamefont {Kim}, \citenamefont {Bertagna}, \citenamefont {D’Souza}, \citenamefont {Heyes}, \citenamefont {Johannissen}, \citenamefont {Nery}, \citenamefont {Pantelias}, \citenamefont {Sanchez-Pedreño~Jimenez}, \citenamefont {Slocombe}, \citenamefont {Spencer}, \citenamefont {Al-Khalili}, \citenamefont {Engel}, \citenamefont {Hay}, \citenamefont {Hingley-Wilson}, \citenamefont {Jeevaratnam}, \citenamefont {Jones}, \citenamefont {Kattnig}, \citenamefont {Lewis}, \citenamefont {Sacchi}, \citenamefont {Scrutton}, \citenamefont {Silva},\ and\ \citenamefont {McFadden}}]{Kim2021}%
  \BibitemOpen
  \bibfield  {author} {\bibinfo {author} {\bibfnamefont {Y.}~\bibnamefont {Kim}}, \bibinfo {author} {\bibfnamefont {F.}~\bibnamefont {Bertagna}}, \bibinfo {author} {\bibfnamefont {E.~M.}\ \bibnamefont {D’Souza}}, \bibinfo {author} {\bibfnamefont {D.~J.}\ \bibnamefont {Heyes}}, \bibinfo {author} {\bibfnamefont {L.~O.}\ \bibnamefont {Johannissen}}, \bibinfo {author} {\bibfnamefont {E.~T.}\ \bibnamefont {Nery}}, \bibinfo {author} {\bibfnamefont {A.}~\bibnamefont {Pantelias}}, \bibinfo {author} {\bibfnamefont {A.}~\bibnamefont {Sanchez-Pedreño~Jimenez}}, \bibinfo {author} {\bibfnamefont {L.}~\bibnamefont {Slocombe}}, \bibinfo {author} {\bibfnamefont {M.~G.}\ \bibnamefont {Spencer}}, \bibinfo {author} {\bibfnamefont {J.}~\bibnamefont {Al-Khalili}}, \bibinfo {author} {\bibfnamefont {G.~S.}\ \bibnamefont {Engel}}, \bibinfo {author} {\bibfnamefont {S.}~\bibnamefont {Hay}}, \bibinfo {author} {\bibfnamefont {S.~M.}\ \bibnamefont {Hingley-Wilson}}, \bibinfo {author} {\bibfnamefont {K.}~\bibnamefont {Jeevaratnam}},
  \bibinfo {author} {\bibfnamefont {A.~R.}\ \bibnamefont {Jones}}, \bibinfo {author} {\bibfnamefont {D.~R.}\ \bibnamefont {Kattnig}}, \bibinfo {author} {\bibfnamefont {R.}~\bibnamefont {Lewis}}, \bibinfo {author} {\bibfnamefont {M.}~\bibnamefont {Sacchi}}, \bibinfo {author} {\bibfnamefont {N.~S.}\ \bibnamefont {Scrutton}}, \bibinfo {author} {\bibfnamefont {S.~R.~P.}\ \bibnamefont {Silva}},\ and\ \bibinfo {author} {\bibfnamefont {J.}~\bibnamefont {McFadden}},\ }\bibfield  {title} {\bibinfo {title} {Quantum biology: {An} update and perspective},\ }\href {https://doi.org/10.3390/quantum3010006} {\bibfield  {journal} {\bibinfo  {journal} {Quantum Rep.}\ }\textbf {\bibinfo {volume} {3}},\ \bibinfo {pages} {80} (\bibinfo {year} {2021})}\BibitemShut {NoStop}%
\bibitem [{\citenamefont {Ritz}\ \emph {et~al.}(2000)\citenamefont {Ritz}, \citenamefont {Adem},\ and\ \citenamefont {Schulten}}]{Ritz2000}%
  \BibitemOpen
  \bibfield  {author} {\bibinfo {author} {\bibfnamefont {T.}~\bibnamefont {Ritz}}, \bibinfo {author} {\bibfnamefont {S.}~\bibnamefont {Adem}},\ and\ \bibinfo {author} {\bibfnamefont {K.}~\bibnamefont {Schulten}},\ }\bibfield  {title} {\bibinfo {title} {A model for photoreceptor-based magnetoreception in birds},\ }\href {https://doi.org/10.1016/S0006-3495(00)76629-X} {\bibfield  {journal} {\bibinfo  {journal} {Biophys. J.}\ }\textbf {\bibinfo {volume} {78}},\ \bibinfo {pages} {707} (\bibinfo {year} {2000})}\BibitemShut {NoStop}%
\bibitem [{\citenamefont {Ramsay}\ \emph {et~al.}(2024)\citenamefont {Ramsay}, \citenamefont {Schuhmann}, \citenamefont {Solov’yov},\ and\ \citenamefont {Kattnig}}]{ramsayCryptochromeMagnetoreceptionTime2024}%
  \BibitemOpen
  \bibfield  {author} {\bibinfo {author} {\bibfnamefont {J.~L.}\ \bibnamefont {Ramsay}}, \bibinfo {author} {\bibfnamefont {F.}~\bibnamefont {Schuhmann}}, \bibinfo {author} {\bibfnamefont {I.~A.}\ \bibnamefont {Solov’yov}},\ and\ \bibinfo {author} {\bibfnamefont {D.~R.}\ \bibnamefont {Kattnig}},\ }\bibfield  {title} {\bibinfo {title} {Cryptochrome magnetoreception: {Time} course of photoactivation from non-equilibrium coarse-grained molecular dynamics},\ }\href {https://doi.org/10.1016/j.csbj.2024.11.001} {\bibfield  {journal} {\bibinfo  {journal} {Comput. Struct. Biotechnol. J.}\ }\textbf {\bibinfo {volume} {26}},\ \bibinfo {pages} {58} (\bibinfo {year} {2024})}\BibitemShut {NoStop}%
\bibitem [{\citenamefont {Frederiksen}\ \emph {et~al.}(2023)\citenamefont {Frederiksen}, \citenamefont {Langebrake}, \citenamefont {Hanić}, \citenamefont {Manthey}, \citenamefont {Mouritsen}, \citenamefont {Liedvogel},\ and\ \citenamefont {Solov’yov}}]{frederiksenMutationalStudyTryptophan2023a}%
  \BibitemOpen
  \bibfield  {author} {\bibinfo {author} {\bibfnamefont {A.}~\bibnamefont {Frederiksen}}, \bibinfo {author} {\bibfnamefont {C.}~\bibnamefont {Langebrake}}, \bibinfo {author} {\bibfnamefont {M.}~\bibnamefont {Hanić}}, \bibinfo {author} {\bibfnamefont {G.}~\bibnamefont {Manthey}}, \bibinfo {author} {\bibfnamefont {H.}~\bibnamefont {Mouritsen}}, \bibinfo {author} {\bibfnamefont {M.}~\bibnamefont {Liedvogel}},\ and\ \bibinfo {author} {\bibfnamefont {I.~A.}\ \bibnamefont {Solov’yov}},\ }\bibfield  {title} {\bibinfo {title} {Mutational {Study} of the {Tryptophan} {Tetrad} {Important} for {Electron} {Transfer} in {European} {Robin} {Cryptochrome} 4a},\ }\href {https://doi.org/10.1021/acsomega.3c02963} {\bibfield  {journal} {\bibinfo  {journal} {ACS Omega}\ }\textbf {\bibinfo {volume} {8}},\ \bibinfo {pages} {26425} (\bibinfo {year} {2023})}\BibitemShut {NoStop}%
\bibitem [{\citenamefont {Wong}\ \emph {et~al.}(2021)\citenamefont {Wong}, \citenamefont {Wei}, \citenamefont {Mouritsen}, \citenamefont {Solov'Yov},\ and\ \citenamefont {Hore}}]{Wong2021}%
  \BibitemOpen
  \bibfield  {author} {\bibinfo {author} {\bibfnamefont {S.~Y.}\ \bibnamefont {Wong}}, \bibinfo {author} {\bibfnamefont {Y.}~\bibnamefont {Wei}}, \bibinfo {author} {\bibfnamefont {H.}~\bibnamefont {Mouritsen}}, \bibinfo {author} {\bibfnamefont {I.~A.}\ \bibnamefont {Solov'Yov}},\ and\ \bibinfo {author} {\bibfnamefont {P.~J.}\ \bibnamefont {Hore}},\ }\bibfield  {title} {\bibinfo {title} {Cryptochrome magnetoreception: four tryptophans could be better than three},\ }\href {https://doi.org/10.1098/rsif.2021.0601} {\bibfield  {journal} {\bibinfo  {journal} {J. R. Soc. Interface}\ }\textbf {\bibinfo {volume} {18}},\ \bibinfo {pages} {20210601} (\bibinfo {year} {2021})}\BibitemShut {NoStop}%
\bibitem [{\citenamefont {Kattnig}\ \emph {et~al.}(2016{\natexlab{a}})\citenamefont {Kattnig}, \citenamefont {Evans}, \citenamefont {Déjean}, \citenamefont {Dodson}, \citenamefont {Wallace}, \citenamefont {Mackenzie}, \citenamefont {Timmel},\ and\ \citenamefont {Hore}}]{Kattnig2016}%
  \BibitemOpen
  \bibfield  {author} {\bibinfo {author} {\bibfnamefont {D.~R.}\ \bibnamefont {Kattnig}}, \bibinfo {author} {\bibfnamefont {E.~W.}\ \bibnamefont {Evans}}, \bibinfo {author} {\bibfnamefont {V.}~\bibnamefont {Déjean}}, \bibinfo {author} {\bibfnamefont {C.~A.}\ \bibnamefont {Dodson}}, \bibinfo {author} {\bibfnamefont {M.~I.}\ \bibnamefont {Wallace}}, \bibinfo {author} {\bibfnamefont {S.~R.}\ \bibnamefont {Mackenzie}}, \bibinfo {author} {\bibfnamefont {C.~R.}\ \bibnamefont {Timmel}},\ and\ \bibinfo {author} {\bibfnamefont {P.~J.}\ \bibnamefont {Hore}},\ }\bibfield  {title} {\bibinfo {title} {Chemical amplification of magnetic field effects relevant to avian magnetoreception},\ }\href {https://doi.org/10.1038/nchem.2447} {\bibfield  {journal} {\bibinfo  {journal} {Nat. Chem}\ }\textbf {\bibinfo {volume} {8}},\ \bibinfo {pages} {384} (\bibinfo {year} {2016}{\natexlab{a}})}\BibitemShut {NoStop}%
\bibitem [{\citenamefont {Nohr}\ \emph {et~al.}(2016)\citenamefont {Nohr}, \citenamefont {Franz}, \citenamefont {Rodriguez}, \citenamefont {Paulus}, \citenamefont {Essen}, \citenamefont {Weber},\ and\ \citenamefont {Schleicher}}]{nohrExtendedElectronTransferAnimal2016}%
  \BibitemOpen
  \bibfield  {author} {\bibinfo {author} {\bibfnamefont {D.}~\bibnamefont {Nohr}}, \bibinfo {author} {\bibfnamefont {S.}~\bibnamefont {Franz}}, \bibinfo {author} {\bibfnamefont {R.}~\bibnamefont {Rodriguez}}, \bibinfo {author} {\bibfnamefont {B.}~\bibnamefont {Paulus}}, \bibinfo {author} {\bibfnamefont {L.-O.}\ \bibnamefont {Essen}}, \bibinfo {author} {\bibfnamefont {S.}~\bibnamefont {Weber}},\ and\ \bibinfo {author} {\bibfnamefont {E.}~\bibnamefont {Schleicher}},\ }\bibfield  {title} {\bibinfo {title} {Extended {Electron}-{Transfer} in {Animal} {Cryptochromes} {Mediated} by a {Tetrad} of {Aromatic} {Amino} {Acids}},\ }\href {https://doi.org/10.1016/j.bpj.2016.06.009} {\bibfield  {journal} {\bibinfo  {journal} {Biophys. J.}\ }\textbf {\bibinfo {volume} {111}},\ \bibinfo {pages} {301} (\bibinfo {year} {2016})}\BibitemShut {NoStop}%
\bibitem [{\citenamefont {Müller}\ \emph {et~al.}(2015)\citenamefont {Müller}, \citenamefont {Yamamoto}, \citenamefont {Martin}, \citenamefont {Iwai},\ and\ \citenamefont {Brettel}}]{mullerDiscoveryFunctionalAnalysis2015}%
  \BibitemOpen
  \bibfield  {author} {\bibinfo {author} {\bibfnamefont {P.}~\bibnamefont {Müller}}, \bibinfo {author} {\bibfnamefont {J.}~\bibnamefont {Yamamoto}}, \bibinfo {author} {\bibfnamefont {R.}~\bibnamefont {Martin}}, \bibinfo {author} {\bibfnamefont {S.}~\bibnamefont {Iwai}},\ and\ \bibinfo {author} {\bibfnamefont {K.}~\bibnamefont {Brettel}},\ }\bibfield  {title} {\bibinfo {title} {Discovery and functional analysis of a 4th electron-transferring tryptophan conserved exclusively in animal cryptochromes and (6-4) photolyases},\ }\href {https://doi.org/10.1039/C5CC06276D} {\bibfield  {journal} {\bibinfo  {journal} {Chem. Commun.}\ }\textbf {\bibinfo {volume} {51}},\ \bibinfo {pages} {15502} (\bibinfo {year} {2015})}\BibitemShut {NoStop}%
\bibitem [{\citenamefont {Liedvogel}\ \emph {et~al.}(2007)\citenamefont {Liedvogel}, \citenamefont {Maeda}, \citenamefont {Henbest}, \citenamefont {Schleicher}, \citenamefont {Simon}, \citenamefont {Timmel}, \citenamefont {Hore},\ and\ \citenamefont {Mouritsen}}]{liedvogelChemicalMagnetoreceptionBird2007}%
  \BibitemOpen
  \bibfield  {author} {\bibinfo {author} {\bibfnamefont {M.}~\bibnamefont {Liedvogel}}, \bibinfo {author} {\bibfnamefont {K.}~\bibnamefont {Maeda}}, \bibinfo {author} {\bibfnamefont {K.}~\bibnamefont {Henbest}}, \bibinfo {author} {\bibfnamefont {E.}~\bibnamefont {Schleicher}}, \bibinfo {author} {\bibfnamefont {T.}~\bibnamefont {Simon}}, \bibinfo {author} {\bibfnamefont {C.~R.}\ \bibnamefont {Timmel}}, \bibinfo {author} {\bibfnamefont {P.~J.}\ \bibnamefont {Hore}},\ and\ \bibinfo {author} {\bibfnamefont {H.}~\bibnamefont {Mouritsen}},\ }\bibfield  {title} {\bibinfo {title} {Chemical magnetoreception: bird cryptochrome 1a is excited by blue light and forms long-lived radical-pairs},\ }\href {https://doi.org/10.1371/journal.pone.0001106} {\bibfield  {journal} {\bibinfo  {journal} {PLoS One}\ }\textbf {\bibinfo {volume} {2}},\ \bibinfo {pages} {e1106} (\bibinfo {year} {2007})}\BibitemShut {NoStop}%
\bibitem [{\citenamefont {Giovani}\ \emph {et~al.}(2003)\citenamefont {Giovani}, \citenamefont {Byrdin}, \citenamefont {Ahmad},\ and\ \citenamefont {Brettel}}]{giovaniLightinducedElectronTransfer2003}%
  \BibitemOpen
  \bibfield  {author} {\bibinfo {author} {\bibfnamefont {B.}~\bibnamefont {Giovani}}, \bibinfo {author} {\bibfnamefont {M.}~\bibnamefont {Byrdin}}, \bibinfo {author} {\bibfnamefont {M.}~\bibnamefont {Ahmad}},\ and\ \bibinfo {author} {\bibfnamefont {K.}~\bibnamefont {Brettel}},\ }\bibfield  {title} {\bibinfo {title} {Light-induced electron transfer in a cryptochrome blue-light photoreceptor},\ }\href {https://doi.org/10.1038/nsb933} {\bibfield  {journal} {\bibinfo  {journal} {Nat. Struct. Mol. Biol.}\ }\textbf {\bibinfo {volume} {10}},\ \bibinfo {pages} {489} (\bibinfo {year} {2003})}\BibitemShut {NoStop}%
\bibitem [{\citenamefont {Deviers}\ \emph {et~al.}(2024)\citenamefont {Deviers}, \citenamefont {Cailliez}, \citenamefont {de~la Lande},\ and\ \citenamefont {Kattnig}}]{deviersAvianCryptochrome42024a}%
  \BibitemOpen
  \bibfield  {author} {\bibinfo {author} {\bibfnamefont {J.}~\bibnamefont {Deviers}}, \bibinfo {author} {\bibfnamefont {F.}~\bibnamefont {Cailliez}}, \bibinfo {author} {\bibfnamefont {A.}~\bibnamefont {de~la Lande}},\ and\ \bibinfo {author} {\bibfnamefont {D.~R.}\ \bibnamefont {Kattnig}},\ }\bibfield  {title} {\bibinfo {title} {Avian cryptochrome 4 binds superoxide},\ }\href {https://doi.org/10.1016/j.csbj.2023.12.009} {\bibfield  {journal} {\bibinfo  {journal} {Comput. Struct. Biotechnol. J.}\ }\textbf {\bibinfo {volume} {26}},\ \bibinfo {pages} {11} (\bibinfo {year} {2024})}\BibitemShut {NoStop}%
\bibitem [{\citenamefont {Salerno}\ \emph {et~al.}(2023)\citenamefont {Salerno}, \citenamefont {Domenico}, \citenamefont {Le}, \citenamefont {Balakrishnan}, \citenamefont {McQuillen}, \citenamefont {Stiles}, \citenamefont {Solov’yov},\ and\ \citenamefont {Martino}}]{salernoLongTimeOxygenSuperoxide2023}%
  \BibitemOpen
  \bibfield  {author} {\bibinfo {author} {\bibfnamefont {K.~M.}\ \bibnamefont {Salerno}}, \bibinfo {author} {\bibfnamefont {J.}~\bibnamefont {Domenico}}, \bibinfo {author} {\bibfnamefont {N.~Q.}\ \bibnamefont {Le}}, \bibinfo {author} {\bibfnamefont {K.}~\bibnamefont {Balakrishnan}}, \bibinfo {author} {\bibfnamefont {R.~J.}\ \bibnamefont {McQuillen}}, \bibinfo {author} {\bibfnamefont {C.~D.}\ \bibnamefont {Stiles}}, \bibinfo {author} {\bibfnamefont {I.~A.}\ \bibnamefont {Solov’yov}},\ and\ \bibinfo {author} {\bibfnamefont {C.~F.}\ \bibnamefont {Martino}},\ }\bibfield  {title} {\bibinfo {title} {Long-{Time} {Oxygen} and {Superoxide} {Localization} in {Arabidopsis} thaliana {Cryptochrome}},\ }\href {https://doi.org/10.1021/acs.jcim.3c00325} {\bibfield  {journal} {\bibinfo  {journal} {J. Chem. Inf. Model.}\ }\textbf {\bibinfo {volume} {63}},\ \bibinfo {pages} {6756} (\bibinfo {year} {2023})}\BibitemShut {NoStop}%
\bibitem [{\citenamefont {Arthaut}\ \emph {et~al.}(2017)\citenamefont {Arthaut}, \citenamefont {Jourdan}, \citenamefont {Mteyrek}, \citenamefont {Procopio}, \citenamefont {El-Esawi}, \citenamefont {d’Harlingue}, \citenamefont {Bouchet}, \citenamefont {Witczak}, \citenamefont {Ritz}, \citenamefont {Klarsfeld}, \citenamefont {Birman}, \citenamefont {Usselman}, \citenamefont {Hoecker}, \citenamefont {Martino},\ and\ \citenamefont {Ahmad}}]{arthautBluelightInducedAccumulation2017}%
  \BibitemOpen
  \bibfield  {author} {\bibinfo {author} {\bibfnamefont {L.-D.}\ \bibnamefont {Arthaut}}, \bibinfo {author} {\bibfnamefont {N.}~\bibnamefont {Jourdan}}, \bibinfo {author} {\bibfnamefont {A.}~\bibnamefont {Mteyrek}}, \bibinfo {author} {\bibfnamefont {M.}~\bibnamefont {Procopio}}, \bibinfo {author} {\bibfnamefont {M.}~\bibnamefont {El-Esawi}}, \bibinfo {author} {\bibfnamefont {A.}~\bibnamefont {d’Harlingue}}, \bibinfo {author} {\bibfnamefont {P.-E.}\ \bibnamefont {Bouchet}}, \bibinfo {author} {\bibfnamefont {J.}~\bibnamefont {Witczak}}, \bibinfo {author} {\bibfnamefont {T.}~\bibnamefont {Ritz}}, \bibinfo {author} {\bibfnamefont {A.}~\bibnamefont {Klarsfeld}}, \bibinfo {author} {\bibfnamefont {S.}~\bibnamefont {Birman}}, \bibinfo {author} {\bibfnamefont {R.~J.}\ \bibnamefont {Usselman}}, \bibinfo {author} {\bibfnamefont {U.}~\bibnamefont {Hoecker}}, \bibinfo {author} {\bibfnamefont {C.~F.}\ \bibnamefont {Martino}},\ and\ \bibinfo {author} {\bibfnamefont {M.}~\bibnamefont {Ahmad}},\ }\bibfield  {title}
  {\bibinfo {title} {Blue-light induced accumulation of reactive oxygen species is a consequence of the {Drosophila} cryptochrome photocycle},\ }\href {https://doi.org/10.1371/journal.pone.0171836} {\bibfield  {journal} {\bibinfo  {journal} {PLOS One}\ }\textbf {\bibinfo {volume} {12}},\ \bibinfo {pages} {e0171836} (\bibinfo {year} {2017})}\BibitemShut {NoStop}%
\bibitem [{\citenamefont {van Wilderen}\ \emph {et~al.}(2015)\citenamefont {van Wilderen}, \citenamefont {Silkstone}, \citenamefont {Mason}, \citenamefont {van Thor},\ and\ \citenamefont {Wilson}}]{vanwilderenKineticStudiesOxidation2015}%
  \BibitemOpen
  \bibfield  {author} {\bibinfo {author} {\bibfnamefont {L.~J. G.~W.}\ \bibnamefont {van Wilderen}}, \bibinfo {author} {\bibfnamefont {G.}~\bibnamefont {Silkstone}}, \bibinfo {author} {\bibfnamefont {M.}~\bibnamefont {Mason}}, \bibinfo {author} {\bibfnamefont {J.~J.}\ \bibnamefont {van Thor}},\ and\ \bibinfo {author} {\bibfnamefont {M.~T.}\ \bibnamefont {Wilson}},\ }\bibfield  {title} {\bibinfo {title} {Kinetic studies on the oxidation of semiquinone and hydroquinone forms of {Arabidopsis} cryptochrome by molecular oxygen},\ }\href {https://doi.org/10.1016/j.fob.2015.10.007} {\bibfield  {journal} {\bibinfo  {journal} {FEBS Open Bio}\ }\textbf {\bibinfo {volume} {5}},\ \bibinfo {pages} {885} (\bibinfo {year} {2015})}\BibitemShut {NoStop}%
\bibitem [{\citenamefont {Müller}\ and\ \citenamefont {Ahmad}(2011)}]{Muller2011}%
  \BibitemOpen
  \bibfield  {author} {\bibinfo {author} {\bibfnamefont {P.}~\bibnamefont {Müller}}\ and\ \bibinfo {author} {\bibfnamefont {M.}~\bibnamefont {Ahmad}},\ }\bibfield  {title} {\bibinfo {title} {Light-activated cryptochrome reacts with molecular oxygen to form a flavin–superoxide radical pair consistent with magnetoreception},\ }\href {https://doi.org/10.1074/JBC.M111.228940} {\bibfield  {journal} {\bibinfo  {journal} {J. Biol. Chem.}\ }\textbf {\bibinfo {volume} {286}},\ \bibinfo {pages} {21033} (\bibinfo {year} {2011})}\BibitemShut {NoStop}%
\bibitem [{\citenamefont {Ritz}\ \emph {et~al.}(2009)\citenamefont {Ritz}, \citenamefont {Wiltschko}, \citenamefont {Hore}, \citenamefont {Rodgers}, \citenamefont {Stapput}, \citenamefont {Thalau}, \citenamefont {Timmel},\ and\ \citenamefont {Wiltschko}}]{Ritz2009}%
  \BibitemOpen
  \bibfield  {author} {\bibinfo {author} {\bibfnamefont {T.}~\bibnamefont {Ritz}}, \bibinfo {author} {\bibfnamefont {R.}~\bibnamefont {Wiltschko}}, \bibinfo {author} {\bibfnamefont {P.~J.}\ \bibnamefont {Hore}}, \bibinfo {author} {\bibfnamefont {C.~T.}\ \bibnamefont {Rodgers}}, \bibinfo {author} {\bibfnamefont {K.}~\bibnamefont {Stapput}}, \bibinfo {author} {\bibfnamefont {P.}~\bibnamefont {Thalau}}, \bibinfo {author} {\bibfnamefont {C.~R.}\ \bibnamefont {Timmel}},\ and\ \bibinfo {author} {\bibfnamefont {W.}~\bibnamefont {Wiltschko}},\ }\bibfield  {title} {\bibinfo {title} {Magnetic compass of birds is based on a molecule with optimal directional sensitivity},\ }\href {https://doi.org/10.1016/J.BPJ.2008.11.072} {\bibfield  {journal} {\bibinfo  {journal} {Biophys. J.}\ }\textbf {\bibinfo {volume} {96}},\ \bibinfo {pages} {3451} (\bibinfo {year} {2009})}\BibitemShut {NoStop}%
\bibitem [{\citenamefont {Massey}(1994)}]{masseyActivationMolecularOxygen1994}%
  \BibitemOpen
  \bibfield  {author} {\bibinfo {author} {\bibfnamefont {V.}~\bibnamefont {Massey}},\ }\bibfield  {title} {\bibinfo {title} {Activation of molecular oxygen by flavins and flavoproteins.},\ }\href {https://doi.org/10.1016/S0021-9258(17)31664-2} {\bibfield  {journal} {\bibinfo  {journal} {J. Biol. Chem.}\ }\textbf {\bibinfo {volume} {269}},\ \bibinfo {pages} {22459} (\bibinfo {year} {1994})}\BibitemShut {NoStop}%
\bibitem [{\citenamefont {Xu}\ \emph {et~al.}(2021)\citenamefont {Xu}, \citenamefont {Jarocha}, \citenamefont {Zollitsch}, \citenamefont {Konowalczyk}, \citenamefont {Henbest}, \citenamefont {Richert}, \citenamefont {Golesworthy}, \citenamefont {Schmidt}, \citenamefont {Déjean}, \citenamefont {Sowood}, \citenamefont {Bassetto}, \citenamefont {Luo}, \citenamefont {Walton}, \citenamefont {Fleming}, \citenamefont {Wei}, \citenamefont {Pitcher}, \citenamefont {Moise}, \citenamefont {Herrmann}, \citenamefont {Yin}, \citenamefont {Wu}, \citenamefont {Bartölke}, \citenamefont {Käsehagen}, \citenamefont {Horst}, \citenamefont {Dautaj}, \citenamefont {Murton}, \citenamefont {Gehrckens}, \citenamefont {Chelliah}, \citenamefont {Takahashi}, \citenamefont {Koch}, \citenamefont {Weber}, \citenamefont {Solov’yov}, \citenamefont {Xie}, \citenamefont {Mackenzie}, \citenamefont {Timmel}, \citenamefont {Mouritsen},\ and\ \citenamefont {Hore}}]{Xu2021}%
  \BibitemOpen
  \bibfield  {author} {\bibinfo {author} {\bibfnamefont {J.}~\bibnamefont {Xu}}, \bibinfo {author} {\bibfnamefont {L.~E.}\ \bibnamefont {Jarocha}}, \bibinfo {author} {\bibfnamefont {T.}~\bibnamefont {Zollitsch}}, \bibinfo {author} {\bibfnamefont {M.}~\bibnamefont {Konowalczyk}}, \bibinfo {author} {\bibfnamefont {K.~B.}\ \bibnamefont {Henbest}}, \bibinfo {author} {\bibfnamefont {S.}~\bibnamefont {Richert}}, \bibinfo {author} {\bibfnamefont {M.~J.}\ \bibnamefont {Golesworthy}}, \bibinfo {author} {\bibfnamefont {J.}~\bibnamefont {Schmidt}}, \bibinfo {author} {\bibfnamefont {V.}~\bibnamefont {Déjean}}, \bibinfo {author} {\bibfnamefont {D.~J.~C.}\ \bibnamefont {Sowood}}, \bibinfo {author} {\bibfnamefont {M.}~\bibnamefont {Bassetto}}, \bibinfo {author} {\bibfnamefont {J.}~\bibnamefont {Luo}}, \bibinfo {author} {\bibfnamefont {J.~R.}\ \bibnamefont {Walton}}, \bibinfo {author} {\bibfnamefont {J.}~\bibnamefont {Fleming}}, \bibinfo {author} {\bibfnamefont {Y.}~\bibnamefont {Wei}}, \bibinfo {author} {\bibfnamefont
  {T.~L.}\ \bibnamefont {Pitcher}}, \bibinfo {author} {\bibfnamefont {G.}~\bibnamefont {Moise}}, \bibinfo {author} {\bibfnamefont {M.}~\bibnamefont {Herrmann}}, \bibinfo {author} {\bibfnamefont {H.}~\bibnamefont {Yin}}, \bibinfo {author} {\bibfnamefont {H.}~\bibnamefont {Wu}}, \bibinfo {author} {\bibfnamefont {R.}~\bibnamefont {Bartölke}}, \bibinfo {author} {\bibfnamefont {S.~J.}\ \bibnamefont {Käsehagen}}, \bibinfo {author} {\bibfnamefont {S.}~\bibnamefont {Horst}}, \bibinfo {author} {\bibfnamefont {G.}~\bibnamefont {Dautaj}}, \bibinfo {author} {\bibfnamefont {P.~D.~F.}\ \bibnamefont {Murton}}, \bibinfo {author} {\bibfnamefont {A.~S.}\ \bibnamefont {Gehrckens}}, \bibinfo {author} {\bibfnamefont {Y.}~\bibnamefont {Chelliah}}, \bibinfo {author} {\bibfnamefont {J.~S.}\ \bibnamefont {Takahashi}}, \bibinfo {author} {\bibfnamefont {K.-W.}\ \bibnamefont {Koch}}, \bibinfo {author} {\bibfnamefont {S.}~\bibnamefont {Weber}}, \bibinfo {author} {\bibfnamefont {I.~A.}\ \bibnamefont {Solov’yov}}, \bibinfo {author}
  {\bibfnamefont {C.}~\bibnamefont {Xie}}, \bibinfo {author} {\bibfnamefont {S.~R.}\ \bibnamefont {Mackenzie}}, \bibinfo {author} {\bibfnamefont {C.~R.}\ \bibnamefont {Timmel}}, \bibinfo {author} {\bibfnamefont {H.}~\bibnamefont {Mouritsen}},\ and\ \bibinfo {author} {\bibfnamefont {P.~J.}\ \bibnamefont {Hore}},\ }\bibfield  {title} {\bibinfo {title} {Magnetic sensitivity of cryptochrome 4 from a migratory songbird},\ }\href {https://doi.org/10.1038/s41586-021-03618-9} {\bibfield  {journal} {\bibinfo  {journal} {Nature}\ }\textbf {\bibinfo {volume} {594}},\ \bibinfo {pages} {535} (\bibinfo {year} {2021})}\BibitemShut {NoStop}%
\bibitem [{\citenamefont {Kerpal}\ \emph {et~al.}(2019)\citenamefont {Kerpal}, \citenamefont {Richert}, \citenamefont {Storey}, \citenamefont {Pillai}, \citenamefont {Liddell}, \citenamefont {Gust}, \citenamefont {Mackenzie}, \citenamefont {Hore},\ and\ \citenamefont {Timmel}}]{Kerpal2019}%
  \BibitemOpen
  \bibfield  {author} {\bibinfo {author} {\bibfnamefont {C.}~\bibnamefont {Kerpal}}, \bibinfo {author} {\bibfnamefont {S.}~\bibnamefont {Richert}}, \bibinfo {author} {\bibfnamefont {J.~G.}\ \bibnamefont {Storey}}, \bibinfo {author} {\bibfnamefont {S.}~\bibnamefont {Pillai}}, \bibinfo {author} {\bibfnamefont {P.~A.}\ \bibnamefont {Liddell}}, \bibinfo {author} {\bibfnamefont {D.}~\bibnamefont {Gust}}, \bibinfo {author} {\bibfnamefont {S.~R.}\ \bibnamefont {Mackenzie}}, \bibinfo {author} {\bibfnamefont {P.~J.}\ \bibnamefont {Hore}},\ and\ \bibinfo {author} {\bibfnamefont {C.~R.}\ \bibnamefont {Timmel}},\ }\bibfield  {title} {\bibinfo {title} {Chemical compass behaviour at microtesla magnetic fields strengthens the radical pair hypothesis of avian magnetoreception},\ }\href {https://doi.org/10.1038/s41467-019-11655-2} {\bibfield  {journal} {\bibinfo  {journal} {Nat. Commun.}\ }\textbf {\bibinfo {volume} {10}},\ \bibinfo {pages} {1} (\bibinfo {year} {2019})}\BibitemShut {NoStop}%
\bibitem [{\citenamefont {Maeda}\ \emph {et~al.}(2008)\citenamefont {Maeda}, \citenamefont {Henbest}, \citenamefont {Cintolesi}, \citenamefont {Kuprov}, \citenamefont {Rodgers}, \citenamefont {Liddell}, \citenamefont {Gust}, \citenamefont {Timmel},\ and\ \citenamefont {Hore}}]{Maeda2008}%
  \BibitemOpen
  \bibfield  {author} {\bibinfo {author} {\bibfnamefont {K.}~\bibnamefont {Maeda}}, \bibinfo {author} {\bibfnamefont {K.~B.}\ \bibnamefont {Henbest}}, \bibinfo {author} {\bibfnamefont {F.}~\bibnamefont {Cintolesi}}, \bibinfo {author} {\bibfnamefont {I.}~\bibnamefont {Kuprov}}, \bibinfo {author} {\bibfnamefont {C.~T.}\ \bibnamefont {Rodgers}}, \bibinfo {author} {\bibfnamefont {P.~A.}\ \bibnamefont {Liddell}}, \bibinfo {author} {\bibfnamefont {D.}~\bibnamefont {Gust}}, \bibinfo {author} {\bibfnamefont {C.~R.}\ \bibnamefont {Timmel}},\ and\ \bibinfo {author} {\bibfnamefont {P.~J.}\ \bibnamefont {Hore}},\ }\bibfield  {title} {\bibinfo {title} {Chemical compass model of avian magnetoreception},\ }\href {https://doi.org/10.1038/nature06834} {\bibfield  {journal} {\bibinfo  {journal} {Nature}\ }\textbf {\bibinfo {volume} {453}},\ \bibinfo {pages} {387} (\bibinfo {year} {2008})}\BibitemShut {NoStop}%
\bibitem [{\citenamefont {Timmel}\ \emph {et~al.}(1998)\citenamefont {Timmel}, \citenamefont {Till}, \citenamefont {Brocklehurst}, \citenamefont {Mclauchlan},\ and\ \citenamefont {Hore}}]{Timmel1998}%
  \BibitemOpen
  \bibfield  {author} {\bibinfo {author} {\bibfnamefont {C.~R.}\ \bibnamefont {Timmel}}, \bibinfo {author} {\bibfnamefont {U.}~\bibnamefont {Till}}, \bibinfo {author} {\bibfnamefont {B.}~\bibnamefont {Brocklehurst}}, \bibinfo {author} {\bibfnamefont {K.~A.}\ \bibnamefont {Mclauchlan}},\ and\ \bibinfo {author} {\bibfnamefont {P.~J.}\ \bibnamefont {Hore}},\ }\bibfield  {title} {\bibinfo {title} {Effects of weak magnetic fields on free radical recombination reactions},\ }\href {https://doi.org/10.1080/00268979809483134} {\bibfield  {journal} {\bibinfo  {journal} {Mol. Phys.}\ }\textbf {\bibinfo {volume} {95}},\ \bibinfo {pages} {71} (\bibinfo {year} {1998})}\BibitemShut {NoStop}%
\bibitem [{\citenamefont {Babcock}\ and\ \citenamefont {Kattnig}(2020)}]{Babcock2020}%
  \BibitemOpen
  \bibfield  {author} {\bibinfo {author} {\bibfnamefont {N.~S.}\ \bibnamefont {Babcock}}\ and\ \bibinfo {author} {\bibfnamefont {D.~R.}\ \bibnamefont {Kattnig}},\ }\bibfield  {title} {\bibinfo {title} {Electron–electron dipolar interaction poses a challenge to the radical pair mechanism of magnetoreception},\ }\href {https://doi.org/10.1021/ACS.JPCLETT.0C00370} {\bibfield  {journal} {\bibinfo  {journal} {J. Phys. Chem. Lett.}\ }\textbf {\bibinfo {volume} {11}},\ \bibinfo {pages} {2414} (\bibinfo {year} {2020})}\BibitemShut {NoStop}%
\bibitem [{\citenamefont {Efimova}\ and\ \citenamefont {Hore}(2008)}]{Efimova2008}%
  \BibitemOpen
  \bibfield  {author} {\bibinfo {author} {\bibfnamefont {O.}~\bibnamefont {Efimova}}\ and\ \bibinfo {author} {\bibfnamefont {P.~J.}\ \bibnamefont {Hore}},\ }\bibfield  {title} {\bibinfo {title} {Role of {Exchange} and {Dipolar} {Interactions} in the {Radical} {Pair} {Model} of the {Avian} {Magnetic} {Compass}},\ }\href {https://doi.org/10.1529/BIOPHYSJ.107.119362} {\bibfield  {journal} {\bibinfo  {journal} {Biophys. J.}\ }\textbf {\bibinfo {volume} {94}},\ \bibinfo {pages} {1565} (\bibinfo {year} {2008})}\BibitemShut {NoStop}%
\bibitem [{\citenamefont {O'Dea}\ \emph {et~al.}(2005)\citenamefont {O'Dea}, \citenamefont {Curtis}, \citenamefont {Green}, \citenamefont {Tinunel},\ and\ \citenamefont {Hore}}]{ODea2005}%
  \BibitemOpen
  \bibfield  {author} {\bibinfo {author} {\bibfnamefont {A.~R.}\ \bibnamefont {O'Dea}}, \bibinfo {author} {\bibfnamefont {A.~F.}\ \bibnamefont {Curtis}}, \bibinfo {author} {\bibfnamefont {N.~J.}\ \bibnamefont {Green}}, \bibinfo {author} {\bibfnamefont {C.~R.}\ \bibnamefont {Tinunel}},\ and\ \bibinfo {author} {\bibfnamefont {P.~J.}\ \bibnamefont {Hore}},\ }\bibfield  {title} {\bibinfo {title} {Influence of {Dipolar} {Interactions} on {Radical} {Pair} {Recombination} {Reactions} {Subject} to {Weak} {Magnetic} {Fields}},\ }\href {https://doi.org/10.1021/JP0456943} {\bibfield  {journal} {\bibinfo  {journal} {J. Phys. Chem. A}\ }\textbf {\bibinfo {volume} {109}},\ \bibinfo {pages} {869} (\bibinfo {year} {2005})}\BibitemShut {NoStop}%
\bibitem [{\citenamefont {Grüning}\ \emph {et~al.}(2024)\citenamefont {Grüning}, \citenamefont {Gerhards}, \citenamefont {Wong}, \citenamefont {Kattnig},\ and\ \citenamefont {Solov'yov}}]{gruningEffectSpinRelaxation2024}%
  \BibitemOpen
  \bibfield  {author} {\bibinfo {author} {\bibfnamefont {G.}~\bibnamefont {Grüning}}, \bibinfo {author} {\bibfnamefont {L.}~\bibnamefont {Gerhards}}, \bibinfo {author} {\bibfnamefont {S.~Y.}\ \bibnamefont {Wong}}, \bibinfo {author} {\bibfnamefont {D.~R.}\ \bibnamefont {Kattnig}},\ and\ \bibinfo {author} {\bibfnamefont {I.~A.}\ \bibnamefont {Solov'yov}},\ }\bibfield  {title} {\bibinfo {title} {The {Effect} of {Spin} {Relaxation} on {Magnetic} {Compass} {Sensitivity} in {ErCry4a}},\ }\href {https://doi.org/10.1002/cphc.202400129} {\bibfield  {journal} {\bibinfo  {journal} {ChemPhysChem}\ }\textbf {\bibinfo {volume} {25}},\ \bibinfo {pages} {e202400129} (\bibinfo {year} {2024})}\BibitemShut {NoStop}%
\bibitem [{\citenamefont {Worster}\ \emph {et~al.}(2016)\citenamefont {Worster}, \citenamefont {Kattnig},\ and\ \citenamefont {Hore}}]{Worster2016}%
  \BibitemOpen
  \bibfield  {author} {\bibinfo {author} {\bibfnamefont {S.}~\bibnamefont {Worster}}, \bibinfo {author} {\bibfnamefont {D.~R.}\ \bibnamefont {Kattnig}},\ and\ \bibinfo {author} {\bibfnamefont {P.~J.}\ \bibnamefont {Hore}},\ }\bibfield  {title} {\bibinfo {title} {Spin relaxation of radicals in cryptochrome and its role in avian magnetoreception},\ }\href {https://doi.org/10.1063/1.4958624} {\bibfield  {journal} {\bibinfo  {journal} {J. Chem. Phys.}\ }\textbf {\bibinfo {volume} {145}},\ \bibinfo {pages} {035104} (\bibinfo {year} {2016})}\BibitemShut {NoStop}%
\bibitem [{\citenamefont {Kattnig}\ \emph {et~al.}(2016{\natexlab{b}})\citenamefont {Kattnig}, \citenamefont {Solov'yov},\ and\ \citenamefont {Hore}}]{Kattnig2016a}%
  \BibitemOpen
  \bibfield  {author} {\bibinfo {author} {\bibfnamefont {D.~R.}\ \bibnamefont {Kattnig}}, \bibinfo {author} {\bibfnamefont {I.~A.}\ \bibnamefont {Solov'yov}},\ and\ \bibinfo {author} {\bibfnamefont {P.~J.}\ \bibnamefont {Hore}},\ }\bibfield  {title} {\bibinfo {title} {Electron spin relaxation in cryptochrome-based magnetoreception},\ }\href {https://doi.org/10.1039/C5CP06731F} {\bibfield  {journal} {\bibinfo  {journal} {Phys. Chem. Chem. Phys.}\ }\textbf {\bibinfo {volume} {18}},\ \bibinfo {pages} {12443} (\bibinfo {year} {2016}{\natexlab{b}})}\BibitemShut {NoStop}%
\bibitem [{\citenamefont {Grüning}\ \emph {et~al.}(2022)\citenamefont {Grüning}, \citenamefont {Wong}, \citenamefont {Gerhards}, \citenamefont {Schuhmann}, \citenamefont {Kattnig}, \citenamefont {Hore},\ and\ \citenamefont {Solov'yov}}]{Gruning2022}%
  \BibitemOpen
  \bibfield  {author} {\bibinfo {author} {\bibfnamefont {G.}~\bibnamefont {Grüning}}, \bibinfo {author} {\bibfnamefont {S.~Y.}\ \bibnamefont {Wong}}, \bibinfo {author} {\bibfnamefont {L.}~\bibnamefont {Gerhards}}, \bibinfo {author} {\bibfnamefont {F.}~\bibnamefont {Schuhmann}}, \bibinfo {author} {\bibfnamefont {D.~R.}\ \bibnamefont {Kattnig}}, \bibinfo {author} {\bibfnamefont {P.~J.}\ \bibnamefont {Hore}},\ and\ \bibinfo {author} {\bibfnamefont {I.~A.}\ \bibnamefont {Solov'yov}},\ }\bibfield  {title} {\bibinfo {title} {Effects of dynamical degrees of freedom on magnetic compass sensitivity: {A} comparison of plant and avian cryptochromes},\ }\href {https://doi.org/10.1021/jacs.2c06233} {\bibfield  {journal} {\bibinfo  {journal} {J. Am. Chem. Soc.}\ }\textbf {\bibinfo {volume} {144}},\ \bibinfo {pages} {22902} (\bibinfo {year} {2022})}\BibitemShut {NoStop}%
\bibitem [{\citenamefont {Smith}\ \emph {et~al.}(2022)\citenamefont {Smith}, \citenamefont {Deviers},\ and\ \citenamefont {Kattnig}}]{Smith2022}%
  \BibitemOpen
  \bibfield  {author} {\bibinfo {author} {\bibfnamefont {L.~D.}\ \bibnamefont {Smith}}, \bibinfo {author} {\bibfnamefont {J.}~\bibnamefont {Deviers}},\ and\ \bibinfo {author} {\bibfnamefont {D.~R.}\ \bibnamefont {Kattnig}},\ }\bibfield  {title} {\bibinfo {title} {Observations about utilitarian coherence in the avian compass},\ }\href {https://doi.org/10.1038/s41598-022-09901-7} {\bibfield  {journal} {\bibinfo  {journal} {Sci. Rep.}\ }\textbf {\bibinfo {volume} {12}},\ \bibinfo {pages} {1} (\bibinfo {year} {2022})}\BibitemShut {NoStop}%
\bibitem [{\citenamefont {Atkins}\ \emph {et~al.}(2019)\citenamefont {Atkins}, \citenamefont {Bajpai}, \citenamefont {Rumball},\ and\ \citenamefont {Kattnig}}]{Atkins2019}%
  \BibitemOpen
  \bibfield  {author} {\bibinfo {author} {\bibfnamefont {C.}~\bibnamefont {Atkins}}, \bibinfo {author} {\bibfnamefont {K.}~\bibnamefont {Bajpai}}, \bibinfo {author} {\bibfnamefont {J.}~\bibnamefont {Rumball}},\ and\ \bibinfo {author} {\bibfnamefont {D.~R.}\ \bibnamefont {Kattnig}},\ }\bibfield  {title} {\bibinfo {title} {On the optimal relative orientation of radicals in the cryptochrome magnetic compass},\ }\href {https://doi.org/10.1063/1.5115445} {\bibfield  {journal} {\bibinfo  {journal} {J. Chem. Phys.}\ }\textbf {\bibinfo {volume} {151}},\ \bibinfo {pages} {065103} (\bibinfo {year} {2019})}\BibitemShut {NoStop}%
\bibitem [{\citenamefont {Smith}\ \emph {et~al.}(2024)\citenamefont {Smith}, \citenamefont {Glatthard}, \citenamefont {Chowdhury},\ and\ \citenamefont {Kattnig}}]{smithOptimalityRadicalpairQuantum2024}%
  \BibitemOpen
  \bibfield  {author} {\bibinfo {author} {\bibfnamefont {L.~D.}\ \bibnamefont {Smith}}, \bibinfo {author} {\bibfnamefont {J.}~\bibnamefont {Glatthard}}, \bibinfo {author} {\bibfnamefont {F.~T.}\ \bibnamefont {Chowdhury}},\ and\ \bibinfo {author} {\bibfnamefont {D.~R.}\ \bibnamefont {Kattnig}},\ }\bibfield  {title} {\bibinfo {title} {On the optimality of the radical-pair quantum compass},\ }\href {https://doi.org/10.1088/2058-9565/ad48b4} {\bibfield  {journal} {\bibinfo  {journal} {Quantum Sci. Technol.}\ }\textbf {\bibinfo {volume} {9}},\ \bibinfo {pages} {035041} (\bibinfo {year} {2024})}\BibitemShut {NoStop}%
\bibitem [{\citenamefont {Procopio}\ and\ \citenamefont {Ritz}(2020)}]{Procopio2020}%
  \BibitemOpen
  \bibfield  {author} {\bibinfo {author} {\bibfnamefont {M.}~\bibnamefont {Procopio}}\ and\ \bibinfo {author} {\bibfnamefont {T.}~\bibnamefont {Ritz}},\ }\bibfield  {title} {\bibinfo {title} {The reference-probe model for a robust and optimal radical-pair-based magnetic compass sensor},\ }\href {https://doi.org/10.1063/1.5128128} {\bibfield  {journal} {\bibinfo  {journal} {J. Chem. Phys.}\ }\textbf {\bibinfo {volume} {152}},\ \bibinfo {pages} {065104} (\bibinfo {year} {2020})}\BibitemShut {NoStop}%
\bibitem [{\citenamefont {Lee}\ \emph {et~al.}(2014)\citenamefont {Lee}, \citenamefont {Lau}, \citenamefont {Hogben}, \citenamefont {Biskup}, \citenamefont {Kattnig},\ and\ \citenamefont {Hore}}]{Lee2014}%
  \BibitemOpen
  \bibfield  {author} {\bibinfo {author} {\bibfnamefont {A.~A.}\ \bibnamefont {Lee}}, \bibinfo {author} {\bibfnamefont {J.~C.~S.}\ \bibnamefont {Lau}}, \bibinfo {author} {\bibfnamefont {H.~J.}\ \bibnamefont {Hogben}}, \bibinfo {author} {\bibfnamefont {T.}~\bibnamefont {Biskup}}, \bibinfo {author} {\bibfnamefont {D.~R.}\ \bibnamefont {Kattnig}},\ and\ \bibinfo {author} {\bibfnamefont {P.~J.}\ \bibnamefont {Hore}},\ }\bibfield  {title} {\bibinfo {title} {Alternative radical pairs for cryptochrome-based magnetoreception},\ }\href {https://doi.org/10.1098/RSIF.2013.1063} {\bibfield  {journal} {\bibinfo  {journal} {J. R. Soc. Interface}\ }\textbf {\bibinfo {volume} {11}},\ \bibinfo {pages} {20131063} (\bibinfo {year} {2014})}\BibitemShut {NoStop}%
\bibitem [{\citenamefont {Player}\ and\ \citenamefont {Hore}(2019)}]{Player2019}%
  \BibitemOpen
  \bibfield  {author} {\bibinfo {author} {\bibfnamefont {T.~C.}\ \bibnamefont {Player}}\ and\ \bibinfo {author} {\bibfnamefont {P.~J.}\ \bibnamefont {Hore}},\ }\bibfield  {title} {\bibinfo {title} {Viability of superoxide-containing radical pairs as magnetoreceptors},\ }\href {https://doi.org/10.1063/1.5129608} {\bibfield  {journal} {\bibinfo  {journal} {J. Chem. Phys.}\ }\textbf {\bibinfo {volume} {151}},\ \bibinfo {pages} {225101} (\bibinfo {year} {2019})}\BibitemShut {NoStop}%
\bibitem [{\citenamefont {Karogodina}\ \emph {et~al.}(2011)\citenamefont {Karogodina}, \citenamefont {Dranov}, \citenamefont {Sergeeva}, \citenamefont {Stass},\ and\ \citenamefont {Steiner}}]{karogodinaKineticMagneticfieldEffect2011}%
  \BibitemOpen
  \bibfield  {author} {\bibinfo {author} {\bibfnamefont {T.~Y.}\ \bibnamefont {Karogodina}}, \bibinfo {author} {\bibfnamefont {I.~G.}\ \bibnamefont {Dranov}}, \bibinfo {author} {\bibfnamefont {S.~V.}\ \bibnamefont {Sergeeva}}, \bibinfo {author} {\bibfnamefont {D.~V.}\ \bibnamefont {Stass}},\ and\ \bibinfo {author} {\bibfnamefont {U.~E.}\ \bibnamefont {Steiner}},\ }\bibfield  {title} {\bibinfo {title} {Kinetic magnetic-field effect involving the small biologically relevant inorganic radicals {NO} and {O2}(·-)},\ }\href {https://doi.org/10.1002/cphc.201100178} {\bibfield  {journal} {\bibinfo  {journal} {ChemPhysChem}\ }\textbf {\bibinfo {volume} {12}},\ \bibinfo {pages} {1714} (\bibinfo {year} {2011})}\BibitemShut {NoStop}%
\bibitem [{\citenamefont {Hogben}\ \emph {et~al.}(2009)\citenamefont {Hogben}, \citenamefont {Efimova}, \citenamefont {Wagner-Rundell}, \citenamefont {Timmel},\ and\ \citenamefont {Hore}}]{Hogben2009}%
  \BibitemOpen
  \bibfield  {author} {\bibinfo {author} {\bibfnamefont {H.~J.}\ \bibnamefont {Hogben}}, \bibinfo {author} {\bibfnamefont {O.}~\bibnamefont {Efimova}}, \bibinfo {author} {\bibfnamefont {N.}~\bibnamefont {Wagner-Rundell}}, \bibinfo {author} {\bibfnamefont {C.~R.}\ \bibnamefont {Timmel}},\ and\ \bibinfo {author} {\bibfnamefont {P.~J.}\ \bibnamefont {Hore}},\ }\bibfield  {title} {\bibinfo {title} {Possible involvement of superoxide and dioxygen with cryptochrome in avian magnetoreception: {Origin} of {Zeeman} resonances observed by in vivo {EPR} spectroscopy},\ }\href {https://doi.org/10.1016/J.CPLETT.2009.08.051} {\bibfield  {journal} {\bibinfo  {journal} {Chem. Phys. Lett.}\ }\textbf {\bibinfo {volume} {480}},\ \bibinfo {pages} {118} (\bibinfo {year} {2009})}\BibitemShut {NoStop}%
\bibitem [{\citenamefont {Mondal}\ and\ \citenamefont {Huix-Rotllant}(2019)}]{mondalTheoreticalInsightsFormation2019}%
  \BibitemOpen
  \bibfield  {author} {\bibinfo {author} {\bibfnamefont {P.}~\bibnamefont {Mondal}}\ and\ \bibinfo {author} {\bibfnamefont {M.}~\bibnamefont {Huix-Rotllant}},\ }\bibfield  {title} {\bibinfo {title} {Theoretical insights into the formation and stability of radical oxygen species in cryptochromes},\ }\href {https://doi.org/10.1039/C9CP00782B} {\bibfield  {journal} {\bibinfo  {journal} {Phys. Chem. Chem. Phys.}\ }\textbf {\bibinfo {volume} {21}},\ \bibinfo {pages} {8874} (\bibinfo {year} {2019})}\BibitemShut {NoStop}%
\bibitem [{\citenamefont {Keens}\ \emph {et~al.}(2018)\citenamefont {Keens}, \citenamefont {Bedkihal},\ and\ \citenamefont {Kattnig}}]{Keens2018}%
  \BibitemOpen
  \bibfield  {author} {\bibinfo {author} {\bibfnamefont {R.~H.}\ \bibnamefont {Keens}}, \bibinfo {author} {\bibfnamefont {S.}~\bibnamefont {Bedkihal}},\ and\ \bibinfo {author} {\bibfnamefont {D.~R.}\ \bibnamefont {Kattnig}},\ }\bibfield  {title} {\bibinfo {title} {Magnetosensitivity in dipolarly coupled three-{Spin} systems},\ }\href {https://doi.org/10.1103/PhysRevLett.121.096001} {\bibfield  {journal} {\bibinfo  {journal} {Phys. Rev. Lett.}\ }\textbf {\bibinfo {volume} {121}},\ \bibinfo {pages} {096001} (\bibinfo {year} {2018})}\BibitemShut {NoStop}%
\bibitem [{\citenamefont {Ramsay}\ and\ \citenamefont {Kattnig}(2022)}]{Ramsay2022}%
  \BibitemOpen
  \bibfield  {author} {\bibinfo {author} {\bibfnamefont {J.}~\bibnamefont {Ramsay}}\ and\ \bibinfo {author} {\bibfnamefont {D.~R.}\ \bibnamefont {Kattnig}},\ }\bibfield  {title} {\bibinfo {title} {Radical triads, not pairs, may explain effects of hypomagnetic fields on neurogenesis},\ }\href {https://doi.org/10.1371/JOURNAL.PCBI.1010519} {\bibfield  {journal} {\bibinfo  {journal} {PLoS Comput. Biol.}\ }\textbf {\bibinfo {volume} {18}},\ \bibinfo {pages} {e1010519} (\bibinfo {year} {2022})}\BibitemShut {NoStop}%
\bibitem [{\citenamefont {Deviers}\ \emph {et~al.}(2022)\citenamefont {Deviers}, \citenamefont {Cailliez}, \citenamefont {De~La~Lande},\ and\ \citenamefont {Kattnig}}]{Deviers2022}%
  \BibitemOpen
  \bibfield  {author} {\bibinfo {author} {\bibfnamefont {J.}~\bibnamefont {Deviers}}, \bibinfo {author} {\bibfnamefont {F.}~\bibnamefont {Cailliez}}, \bibinfo {author} {\bibfnamefont {A.}~\bibnamefont {De~La~Lande}},\ and\ \bibinfo {author} {\bibfnamefont {D.~R.}\ \bibnamefont {Kattnig}},\ }\bibfield  {title} {\bibinfo {title} {Anisotropic magnetic field effects in the re-oxidation of cryptochrome in the presence of scavenger radicals},\ }\href {https://doi.org/10.1063/5.0078115} {\bibfield  {journal} {\bibinfo  {journal} {J. Chem. Phys.}\ }\textbf {\bibinfo {volume} {156}},\ \bibinfo {pages} {025101} (\bibinfo {year} {2022})}\BibitemShut {NoStop}%
\bibitem [{\citenamefont {Babcock}\ and\ \citenamefont {Kattnig}(2021)}]{Babcock2021}%
  \BibitemOpen
  \bibfield  {author} {\bibinfo {author} {\bibfnamefont {N.~S.}\ \bibnamefont {Babcock}}\ and\ \bibinfo {author} {\bibfnamefont {D.~R.}\ \bibnamefont {Kattnig}},\ }\bibfield  {title} {\bibinfo {title} {Radical scavenging could answer the challenge posed by electron–electron dipolar interactions in the cryptochrome compass model},\ }\href {https://doi.org/10.1021/jacsau.1c00332} {\bibfield  {journal} {\bibinfo  {journal} {JACS Au}\ }\textbf {\bibinfo {volume} {1}},\ \bibinfo {pages} {2033} (\bibinfo {year} {2021})}\BibitemShut {NoStop}%
\bibitem [{\citenamefont {Kattnig}\ and\ \citenamefont {Hore}(2017)}]{Kattnig2017}%
  \BibitemOpen
  \bibfield  {author} {\bibinfo {author} {\bibfnamefont {D.~R.}\ \bibnamefont {Kattnig}}\ and\ \bibinfo {author} {\bibfnamefont {P.~J.}\ \bibnamefont {Hore}},\ }\bibfield  {title} {\bibinfo {title} {The sensitivity of a radical pair compass magnetoreceptor can be significantly amplified by radical scavengers},\ }\href {https://doi.org/10.1038/s41598-017-09914-7} {\bibfield  {journal} {\bibinfo  {journal} {Sci. Rep.}\ }\textbf {\bibinfo {volume} {7}},\ \bibinfo {pages} {1} (\bibinfo {year} {2017})}\BibitemShut {NoStop}%
\bibitem [{\citenamefont {Luo}(2024)}]{luoSensitivityEnhancementRadicalpair2024a}%
  \BibitemOpen
  \bibfield  {author} {\bibinfo {author} {\bibfnamefont {J.}~\bibnamefont {Luo}},\ }\bibfield  {title} {\bibinfo {title} {Sensitivity enhancement of radical-pair magnetoreceptors as a result of spin decoherence},\ }\href {https://doi.org/10.1063/5.0182172} {\bibfield  {journal} {\bibinfo  {journal} {J. Chem. Phys.}\ }\textbf {\bibinfo {volume} {160}},\ \bibinfo {pages} {074306} (\bibinfo {year} {2024})}\BibitemShut {NoStop}%
\bibitem [{\citenamefont {Ramsay}\ and\ \citenamefont {Kattnig}(2023)}]{Ramsay2023}%
  \BibitemOpen
  \bibfield  {author} {\bibinfo {author} {\bibfnamefont {J.~L.}\ \bibnamefont {Ramsay}}\ and\ \bibinfo {author} {\bibfnamefont {D.~R.}\ \bibnamefont {Kattnig}},\ }\bibfield  {title} {\bibinfo {title} {Magnetoreception in cryptochrome enabled by one-dimensional radical motion},\ }\href {https://doi.org/10.1116/5.0142227} {\bibfield  {journal} {\bibinfo  {journal} {AVS Quantum Sci.}\ }\textbf {\bibinfo {volume} {5}},\ \bibinfo {pages} {22601} (\bibinfo {year} {2023})}\BibitemShut {NoStop}%
\bibitem [{\citenamefont {Burgarth}\ \emph {et~al.}(2020)\citenamefont {Burgarth}, \citenamefont {Facchi}, \citenamefont {Nakazato}, \citenamefont {Pascazio},\ and\ \citenamefont {Yuasa}}]{burgarthQuantumZenoDynamics2020}%
  \BibitemOpen
  \bibfield  {author} {\bibinfo {author} {\bibfnamefont {D.}~\bibnamefont {Burgarth}}, \bibinfo {author} {\bibfnamefont {P.}~\bibnamefont {Facchi}}, \bibinfo {author} {\bibfnamefont {H.}~\bibnamefont {Nakazato}}, \bibinfo {author} {\bibfnamefont {S.}~\bibnamefont {Pascazio}},\ and\ \bibinfo {author} {\bibfnamefont {K.}~\bibnamefont {Yuasa}},\ }\bibfield  {title} {\bibinfo {title} {Quantum {Zeno} {Dynamics} from {General} {Quantum} {Operations}},\ }\href {https://doi.org/10.22331/q-2020-07-06-289} {\bibfield  {journal} {\bibinfo  {journal} {Quantum}\ }\textbf {\bibinfo {volume} {4}},\ \bibinfo {pages} {289} (\bibinfo {year} {2020})}\BibitemShut {NoStop}%
\bibitem [{\citenamefont {Dellis}\ and\ \citenamefont {Kominis}(2012)}]{Dellis2012}%
  \BibitemOpen
  \bibfield  {author} {\bibinfo {author} {\bibfnamefont {A.~T.}\ \bibnamefont {Dellis}}\ and\ \bibinfo {author} {\bibfnamefont {I.~K.}\ \bibnamefont {Kominis}},\ }\bibfield  {title} {\bibinfo {title} {The quantum {Zeno} effect immunizes the avian compass against the deleterious effects of exchange and dipolar interactions},\ }\href {https://doi.org/10.1016/J.BIOSYSTEMS.2011.11.007} {\bibfield  {journal} {\bibinfo  {journal} {Biosyst.}\ }\textbf {\bibinfo {volume} {107}},\ \bibinfo {pages} {153} (\bibinfo {year} {2012})}\BibitemShut {NoStop}%
\bibitem [{\citenamefont {Berdinskii}\ and\ \citenamefont {Yakunin}(2008)}]{berdinskiiChemicalZenoEffect2008}%
  \BibitemOpen
  \bibfield  {author} {\bibinfo {author} {\bibfnamefont {V.~L.}\ \bibnamefont {Berdinskii}}\ and\ \bibinfo {author} {\bibfnamefont {I.~N.}\ \bibnamefont {Yakunin}},\ }\bibfield  {title} {\bibinfo {title} {Chemical {Zeno} effect and its manifestations},\ }\href {https://doi.org/10.1134/S0012501608070026} {\bibfield  {journal} {\bibinfo  {journal} {Dokl. Phys. Chem.}\ }\textbf {\bibinfo {volume} {421}},\ \bibinfo {pages} {163} (\bibinfo {year} {2008})}\BibitemShut {NoStop}%
\bibitem [{\citenamefont {Itano}\ \emph {et~al.}(1990)\citenamefont {Itano}, \citenamefont {Heinzen}, \citenamefont {Bollinger},\ and\ \citenamefont {Wineland}}]{itanoQuantumZenoEffect1990}%
  \BibitemOpen
  \bibfield  {author} {\bibinfo {author} {\bibfnamefont {W.~M.}\ \bibnamefont {Itano}}, \bibinfo {author} {\bibfnamefont {D.~J.}\ \bibnamefont {Heinzen}}, \bibinfo {author} {\bibfnamefont {J.~J.}\ \bibnamefont {Bollinger}},\ and\ \bibinfo {author} {\bibfnamefont {D.~J.}\ \bibnamefont {Wineland}},\ }\bibfield  {title} {\bibinfo {title} {Quantum {Zeno} effect},\ }\href {https://doi.org/10.1103/PhysRevA.41.2295} {\bibfield  {journal} {\bibinfo  {journal} {Phys. Rev. A}\ }\textbf {\bibinfo {volume} {41}},\ \bibinfo {pages} {2295} (\bibinfo {year} {1990})}\BibitemShut {NoStop}%
\bibitem [{\citenamefont {Misra}\ and\ \citenamefont {Sudarshan}(1977)}]{misraZenosParadoxQuantum1977}%
  \BibitemOpen
  \bibfield  {author} {\bibinfo {author} {\bibfnamefont {B.}~\bibnamefont {Misra}}\ and\ \bibinfo {author} {\bibfnamefont {E.~C.~G.}\ \bibnamefont {Sudarshan}},\ }\bibfield  {title} {\bibinfo {title} {The {Zeno}’s paradox in quantum theory},\ }\href {https://doi.org/10.1063/1.523304} {\bibfield  {journal} {\bibinfo  {journal} {J. Math. Phys.}\ }\textbf {\bibinfo {volume} {18}},\ \bibinfo {pages} {756} (\bibinfo {year} {1977})}\BibitemShut {NoStop}%
\bibitem [{\citenamefont {Denton}\ \emph {et~al.}(2024)\citenamefont {Denton}, \citenamefont {Smith}, \citenamefont {Xu}, \citenamefont {Pugsley}, \citenamefont {Toghill},\ and\ \citenamefont {Kattnig}}]{dentonMagnetosensitivityTightlyBound2024}%
  \BibitemOpen
  \bibfield  {author} {\bibinfo {author} {\bibfnamefont {M.~C.~J.}\ \bibnamefont {Denton}}, \bibinfo {author} {\bibfnamefont {L.~D.}\ \bibnamefont {Smith}}, \bibinfo {author} {\bibfnamefont {W.}~\bibnamefont {Xu}}, \bibinfo {author} {\bibfnamefont {J.}~\bibnamefont {Pugsley}}, \bibinfo {author} {\bibfnamefont {A.}~\bibnamefont {Toghill}},\ and\ \bibinfo {author} {\bibfnamefont {D.~R.}\ \bibnamefont {Kattnig}},\ }\bibfield  {title} {\bibinfo {title} {Magnetosensitivity of tightly bound radical pairs in cryptochrome is enabled by the quantum {Zeno} effect},\ }\href {https://doi.org/10.1038/s41467-024-55124-x} {\bibfield  {journal} {\bibinfo  {journal} {Nat. Commun.}\ }\textbf {\bibinfo {volume} {15}},\ \bibinfo {pages} {10823} (\bibinfo {year} {2024})}\BibitemShut {NoStop}%
\bibitem [{\citenamefont {Chiesa}\ \emph {et~al.}(2023)\citenamefont {Chiesa}, \citenamefont {Privitera}, \citenamefont {Macaluso}, \citenamefont {Mannini}, \citenamefont {Bittl}, \citenamefont {Naaman}, \citenamefont {Wasielewski}, \citenamefont {Sessoli},\ and\ \citenamefont {Carretta}}]{chiesaChiralityInducedSpinSelectivity2023}%
  \BibitemOpen
  \bibfield  {author} {\bibinfo {author} {\bibfnamefont {A.}~\bibnamefont {Chiesa}}, \bibinfo {author} {\bibfnamefont {A.}~\bibnamefont {Privitera}}, \bibinfo {author} {\bibfnamefont {E.}~\bibnamefont {Macaluso}}, \bibinfo {author} {\bibfnamefont {M.}~\bibnamefont {Mannini}}, \bibinfo {author} {\bibfnamefont {R.}~\bibnamefont {Bittl}}, \bibinfo {author} {\bibfnamefont {R.}~\bibnamefont {Naaman}}, \bibinfo {author} {\bibfnamefont {M.~R.}\ \bibnamefont {Wasielewski}}, \bibinfo {author} {\bibfnamefont {R.}~\bibnamefont {Sessoli}},\ and\ \bibinfo {author} {\bibfnamefont {S.}~\bibnamefont {Carretta}},\ }\bibfield  {title} {\bibinfo {title} {Chirality-{Induced} {Spin} {Selectivity}: {An} {Enabling} {Technology} for {Quantum} {Applications}},\ }\href {https://doi.org/10.1002/adma.202300472} {\bibfield  {journal} {\bibinfo  {journal} {Adv. Mater.}\ }\textbf {\bibinfo {volume} {35}},\ \bibinfo {pages} {2300472} (\bibinfo {year} {2023})}\BibitemShut {NoStop}%
\bibitem [{\citenamefont {Aiello}\ \emph {et~al.}(2022)\citenamefont {Aiello}, \citenamefont {Abendroth}, \citenamefont {Abbas}, \citenamefont {Afanasev}, \citenamefont {Agarwal}, \citenamefont {Banerjee}, \citenamefont {Beratan}, \citenamefont {Belling}, \citenamefont {Berche}, \citenamefont {Botana}, \citenamefont {Caram}, \citenamefont {Celardo}, \citenamefont {Cuniberti}, \citenamefont {Garcia-Etxarri}, \citenamefont {Dianat}, \citenamefont {Diez-Perez}, \citenamefont {Guo}, \citenamefont {Gutierrez}, \citenamefont {Herrmann}, \citenamefont {Hihath}, \citenamefont {Kale}, \citenamefont {Kurian}, \citenamefont {Lai}, \citenamefont {Liu}, \citenamefont {Lopez}, \citenamefont {Medina}, \citenamefont {Mujica}, \citenamefont {Naaman}, \citenamefont {Noormandipour}, \citenamefont {Palma}, \citenamefont {Paltiel}, \citenamefont {Petuskey}, \citenamefont {Ribeiro-Silva}, \citenamefont {Saenz}, \citenamefont {Santos}, \citenamefont {Solyanik-Gorgone}, \citenamefont {Sorger}, \citenamefont {Stemer}, \citenamefont
  {Ugalde}, \citenamefont {Valdes-Curiel}, \citenamefont {Varela}, \citenamefont {Waldeck}, \citenamefont {Wasielewski}, \citenamefont {Weiss}, \citenamefont {Zacharias},\ and\ \citenamefont {Wang}}]{aielloChiralityBasedQuantumLeap2022}%
  \BibitemOpen
  \bibfield  {author} {\bibinfo {author} {\bibfnamefont {C.~D.}\ \bibnamefont {Aiello}}, \bibinfo {author} {\bibfnamefont {J.~M.}\ \bibnamefont {Abendroth}}, \bibinfo {author} {\bibfnamefont {M.}~\bibnamefont {Abbas}}, \bibinfo {author} {\bibfnamefont {A.}~\bibnamefont {Afanasev}}, \bibinfo {author} {\bibfnamefont {S.}~\bibnamefont {Agarwal}}, \bibinfo {author} {\bibfnamefont {A.~S.}\ \bibnamefont {Banerjee}}, \bibinfo {author} {\bibfnamefont {D.~N.}\ \bibnamefont {Beratan}}, \bibinfo {author} {\bibfnamefont {J.~N.}\ \bibnamefont {Belling}}, \bibinfo {author} {\bibfnamefont {B.}~\bibnamefont {Berche}}, \bibinfo {author} {\bibfnamefont {A.}~\bibnamefont {Botana}}, \bibinfo {author} {\bibfnamefont {J.~R.}\ \bibnamefont {Caram}}, \bibinfo {author} {\bibfnamefont {G.~L.}\ \bibnamefont {Celardo}}, \bibinfo {author} {\bibfnamefont {G.}~\bibnamefont {Cuniberti}}, \bibinfo {author} {\bibfnamefont {A.}~\bibnamefont {Garcia-Etxarri}}, \bibinfo {author} {\bibfnamefont {A.}~\bibnamefont {Dianat}}, \bibinfo {author}
  {\bibfnamefont {I.}~\bibnamefont {Diez-Perez}}, \bibinfo {author} {\bibfnamefont {Y.}~\bibnamefont {Guo}}, \bibinfo {author} {\bibfnamefont {R.}~\bibnamefont {Gutierrez}}, \bibinfo {author} {\bibfnamefont {C.}~\bibnamefont {Herrmann}}, \bibinfo {author} {\bibfnamefont {J.}~\bibnamefont {Hihath}}, \bibinfo {author} {\bibfnamefont {S.}~\bibnamefont {Kale}}, \bibinfo {author} {\bibfnamefont {P.}~\bibnamefont {Kurian}}, \bibinfo {author} {\bibfnamefont {Y.-C.}\ \bibnamefont {Lai}}, \bibinfo {author} {\bibfnamefont {T.}~\bibnamefont {Liu}}, \bibinfo {author} {\bibfnamefont {A.}~\bibnamefont {Lopez}}, \bibinfo {author} {\bibfnamefont {E.}~\bibnamefont {Medina}}, \bibinfo {author} {\bibfnamefont {V.}~\bibnamefont {Mujica}}, \bibinfo {author} {\bibfnamefont {R.}~\bibnamefont {Naaman}}, \bibinfo {author} {\bibfnamefont {M.}~\bibnamefont {Noormandipour}}, \bibinfo {author} {\bibfnamefont {J.~L.}\ \bibnamefont {Palma}}, \bibinfo {author} {\bibfnamefont {Y.}~\bibnamefont {Paltiel}}, \bibinfo {author} {\bibfnamefont
  {W.}~\bibnamefont {Petuskey}}, \bibinfo {author} {\bibfnamefont {J.~C.}\ \bibnamefont {Ribeiro-Silva}}, \bibinfo {author} {\bibfnamefont {J.~J.}\ \bibnamefont {Saenz}}, \bibinfo {author} {\bibfnamefont {E.~J.~G.}\ \bibnamefont {Santos}}, \bibinfo {author} {\bibfnamefont {M.}~\bibnamefont {Solyanik-Gorgone}}, \bibinfo {author} {\bibfnamefont {V.~J.}\ \bibnamefont {Sorger}}, \bibinfo {author} {\bibfnamefont {D.~M.}\ \bibnamefont {Stemer}}, \bibinfo {author} {\bibfnamefont {J.~M.}\ \bibnamefont {Ugalde}}, \bibinfo {author} {\bibfnamefont {A.}~\bibnamefont {Valdes-Curiel}}, \bibinfo {author} {\bibfnamefont {S.}~\bibnamefont {Varela}}, \bibinfo {author} {\bibfnamefont {D.~H.}\ \bibnamefont {Waldeck}}, \bibinfo {author} {\bibfnamefont {M.~R.}\ \bibnamefont {Wasielewski}}, \bibinfo {author} {\bibfnamefont {P.~S.}\ \bibnamefont {Weiss}}, \bibinfo {author} {\bibfnamefont {H.}~\bibnamefont {Zacharias}},\ and\ \bibinfo {author} {\bibfnamefont {Q.~H.}\ \bibnamefont {Wang}},\ }\bibfield  {title} {\bibinfo {title} {A
  {Chirality}-{Based} {Quantum} {Leap}},\ }\href {https://doi.org/10.1021/acsnano.1c01347} {\bibfield  {journal} {\bibinfo  {journal} {ACS Nano}\ }\textbf {\bibinfo {volume} {16}},\ \bibinfo {pages} {4989} (\bibinfo {year} {2022})}\BibitemShut {NoStop}%
\bibitem [{\citenamefont {Abendroth}\ \emph {et~al.}(2019)\citenamefont {Abendroth}, \citenamefont {Stemer}, \citenamefont {Bloom}, \citenamefont {Roy}, \citenamefont {Naaman}, \citenamefont {Waldeck}, \citenamefont {Weiss},\ and\ \citenamefont {Mondal}}]{abendrothSpinSelectivityPhotoinduced2019}%
  \BibitemOpen
  \bibfield  {author} {\bibinfo {author} {\bibfnamefont {J.~M.}\ \bibnamefont {Abendroth}}, \bibinfo {author} {\bibfnamefont {D.~M.}\ \bibnamefont {Stemer}}, \bibinfo {author} {\bibfnamefont {B.~P.}\ \bibnamefont {Bloom}}, \bibinfo {author} {\bibfnamefont {P.}~\bibnamefont {Roy}}, \bibinfo {author} {\bibfnamefont {R.}~\bibnamefont {Naaman}}, \bibinfo {author} {\bibfnamefont {D.~H.}\ \bibnamefont {Waldeck}}, \bibinfo {author} {\bibfnamefont {P.~S.}\ \bibnamefont {Weiss}},\ and\ \bibinfo {author} {\bibfnamefont {P.~C.}\ \bibnamefont {Mondal}},\ }\bibfield  {title} {\bibinfo {title} {Spin {Selectivity} in {Photoinduced} {Charge}-{Transfer} {Mediated} by {Chiral} {Molecules}},\ }\href {https://doi.org/10.1021/acsnano.9b01876} {\bibfield  {journal} {\bibinfo  {journal} {ACS Nano}\ }\textbf {\bibinfo {volume} {13}},\ \bibinfo {pages} {4928} (\bibinfo {year} {2019})}\BibitemShut {NoStop}%
\bibitem [{\citenamefont {Bloom}\ \emph {et~al.}(2017)\citenamefont {Bloom}, \citenamefont {Graff}, \citenamefont {Ghosh}, \citenamefont {Beratan},\ and\ \citenamefont {Waldeck}}]{bloomChiralityControlElectron2017}%
  \BibitemOpen
  \bibfield  {author} {\bibinfo {author} {\bibfnamefont {B.~P.}\ \bibnamefont {Bloom}}, \bibinfo {author} {\bibfnamefont {B.~M.}\ \bibnamefont {Graff}}, \bibinfo {author} {\bibfnamefont {S.}~\bibnamefont {Ghosh}}, \bibinfo {author} {\bibfnamefont {D.~N.}\ \bibnamefont {Beratan}},\ and\ \bibinfo {author} {\bibfnamefont {D.~H.}\ \bibnamefont {Waldeck}},\ }\bibfield  {title} {\bibinfo {title} {Chirality {Control} of {Electron} {Transfer} in {Quantum} {Dot} {Assemblies}},\ }\href {https://doi.org/10.1021/jacs.7b04639} {\bibfield  {journal} {\bibinfo  {journal} {J. Am. Chem. Soc.}\ }\textbf {\bibinfo {volume} {139}},\ \bibinfo {pages} {9038} (\bibinfo {year} {2017})}\BibitemShut {NoStop}%
\bibitem [{\citenamefont {Michaeli}\ \emph {et~al.}(2016)\citenamefont {Michaeli}, \citenamefont {Kantor-Uriel}, \citenamefont {Naaman},\ and\ \citenamefont {Waldeck}}]{michaeliElectronsSpinMolecular2016}%
  \BibitemOpen
  \bibfield  {author} {\bibinfo {author} {\bibfnamefont {K.}~\bibnamefont {Michaeli}}, \bibinfo {author} {\bibfnamefont {N.}~\bibnamefont {Kantor-Uriel}}, \bibinfo {author} {\bibfnamefont {R.}~\bibnamefont {Naaman}},\ and\ \bibinfo {author} {\bibfnamefont {D.~H.}\ \bibnamefont {Waldeck}},\ }\bibfield  {title} {\bibinfo {title} {The electron's spin and molecular chirality – how are they related and how do they affect life processes?},\ }\href {https://doi.org/10.1039/C6CS00369A} {\bibfield  {journal} {\bibinfo  {journal} {Chem. Soc. Rev.}\ }\textbf {\bibinfo {volume} {45}},\ \bibinfo {pages} {6478} (\bibinfo {year} {2016})}\BibitemShut {NoStop}%
\bibitem [{\citenamefont {Wasielewski}(2006)}]{wasielewskiEnergyChargeSpin2006}%
  \BibitemOpen
  \bibfield  {author} {\bibinfo {author} {\bibfnamefont {M.~R.}\ \bibnamefont {Wasielewski}},\ }\bibfield  {title} {\bibinfo {title} {Energy, {Charge}, and {Spin} {Transport} in {Molecules} and {Self}-{Assembled} {Nanostructures} {Inspired} by {Photosynthesis}},\ }\href {https://doi.org/10.1021/jo060225d} {\bibfield  {journal} {\bibinfo  {journal} {J. Org. Chem.}\ }\textbf {\bibinfo {volume} {71}},\ \bibinfo {pages} {5051} (\bibinfo {year} {2006})}\BibitemShut {NoStop}%
\bibitem [{\citenamefont {Fay}(2021)}]{fayChiralityInducedSpinCoherence2021}%
  \BibitemOpen
  \bibfield  {author} {\bibinfo {author} {\bibfnamefont {T.~P.}\ \bibnamefont {Fay}},\ }\bibfield  {title} {\bibinfo {title} {Chirality-{Induced} {Spin} {Coherence} in {Electron} {Transfer} {Reactions}},\ }\href {https://doi.org/10.1021/acs.jpclett.1c00009} {\bibfield  {journal} {\bibinfo  {journal} {J. Phys. Chem. Lett.}\ }\textbf {\bibinfo {volume} {12}},\ \bibinfo {pages} {1407} (\bibinfo {year} {2021})}\BibitemShut {NoStop}%
\bibitem [{\citenamefont {Fay}\ and\ \citenamefont {Limmer}(2023)}]{faySpinSelectiveCharge2023}%
  \BibitemOpen
  \bibfield  {author} {\bibinfo {author} {\bibfnamefont {T.~P.}\ \bibnamefont {Fay}}\ and\ \bibinfo {author} {\bibfnamefont {D.~T.}\ \bibnamefont {Limmer}},\ }\bibfield  {title} {\bibinfo {title} {Spin selective charge recombination in chiral donor–bridge–acceptor triads},\ }\href {https://doi.org/10.1063/5.0150269} {\bibfield  {journal} {\bibinfo  {journal} {J. Chem. Phys.}\ }\textbf {\bibinfo {volume} {158}},\ \bibinfo {pages} {194101} (\bibinfo {year} {2023})}\BibitemShut {NoStop}%
\bibitem [{\citenamefont {Vittmann}\ \emph {et~al.}(2023)\citenamefont {Vittmann}, \citenamefont {Lim}, \citenamefont {Tamascelli}, \citenamefont {Huelga},\ and\ \citenamefont {Plenio}}]{vittmannSpinDependentMomentumConservation2023}%
  \BibitemOpen
  \bibfield  {author} {\bibinfo {author} {\bibfnamefont {C.}~\bibnamefont {Vittmann}}, \bibinfo {author} {\bibfnamefont {J.}~\bibnamefont {Lim}}, \bibinfo {author} {\bibfnamefont {D.}~\bibnamefont {Tamascelli}}, \bibinfo {author} {\bibfnamefont {S.~F.}\ \bibnamefont {Huelga}},\ and\ \bibinfo {author} {\bibfnamefont {M.~B.}\ \bibnamefont {Plenio}},\ }\bibfield  {title} {\bibinfo {title} {Spin-{Dependent} {Momentum} {Conservation} of {Electron}–{Phonon} {Scattering} in {Chirality}-{Induced} {Spin} {Selectivity}},\ }\href {https://doi.org/10.1021/acs.jpclett.2c03224} {\bibfield  {journal} {\bibinfo  {journal} {J. Phys. Chem. Lett.}\ }\textbf {\bibinfo {volume} {14}},\ \bibinfo {pages} {340} (\bibinfo {year} {2023})}\BibitemShut {NoStop}%
\bibitem [{\citenamefont {Vittmann}\ \emph {et~al.}(2022)\citenamefont {Vittmann}, \citenamefont {Kessing}, \citenamefont {Lim}, \citenamefont {Huelga},\ and\ \citenamefont {Plenio}}]{vittmannInterfaceInducedConservationMomentum2022}%
  \BibitemOpen
  \bibfield  {author} {\bibinfo {author} {\bibfnamefont {C.}~\bibnamefont {Vittmann}}, \bibinfo {author} {\bibfnamefont {R.~K.}\ \bibnamefont {Kessing}}, \bibinfo {author} {\bibfnamefont {J.}~\bibnamefont {Lim}}, \bibinfo {author} {\bibfnamefont {S.~F.}\ \bibnamefont {Huelga}},\ and\ \bibinfo {author} {\bibfnamefont {M.~B.}\ \bibnamefont {Plenio}},\ }\bibfield  {title} {\bibinfo {title} {Interface-{Induced} {Conservation} of {Momentum} {Leads} to {Chiral}-{Induced} {Spin} {Selectivity}},\ }\href {https://doi.org/10.1021/acs.jpclett.1c03975} {\bibfield  {journal} {\bibinfo  {journal} {J. Phys. Chem. Lett.}\ }\textbf {\bibinfo {volume} {13}},\ \bibinfo {pages} {1791} (\bibinfo {year} {2022})}\BibitemShut {NoStop}%
\bibitem [{\citenamefont {Fay}\ and\ \citenamefont {Limmer}(2021)}]{fayOriginChiralityInduced2021}%
  \BibitemOpen
  \bibfield  {author} {\bibinfo {author} {\bibfnamefont {T.~P.}\ \bibnamefont {Fay}}\ and\ \bibinfo {author} {\bibfnamefont {D.~T.}\ \bibnamefont {Limmer}},\ }\bibfield  {title} {\bibinfo {title} {Origin of {Chirality} {Induced} {Spin} {Selectivity} in {Photoinduced} {Electron} {Transfer}},\ }\href {https://doi.org/10.1021/acs.nanolett.1c02370} {\bibfield  {journal} {\bibinfo  {journal} {Nano Lett.}\ }\textbf {\bibinfo {volume} {21}},\ \bibinfo {pages} {6696} (\bibinfo {year} {2021})}\BibitemShut {NoStop}%
\bibitem [{\citenamefont {Luo}\ and\ \citenamefont {Hore}(2021)}]{Luo2021}%
  \BibitemOpen
  \bibfield  {author} {\bibinfo {author} {\bibfnamefont {J.}~\bibnamefont {Luo}}\ and\ \bibinfo {author} {\bibfnamefont {P.~J.}\ \bibnamefont {Hore}},\ }\bibfield  {title} {\bibinfo {title} {Chiral-induced spin selectivity in the formation and recombination of radical pairs: {Cryptochrome} magnetoreception and {EPR} detection},\ }\href {https://doi.org/10.1088/1367-2630/ABED0B} {\bibfield  {journal} {\bibinfo  {journal} {New J. Phys.}\ }\textbf {\bibinfo {volume} {23}},\ \bibinfo {pages} {043032} (\bibinfo {year} {2021})}\BibitemShut {NoStop}%
\bibitem [{\citenamefont {Chiesa}\ \emph {et~al.}(2021)\citenamefont {Chiesa}, \citenamefont {Chizzini}, \citenamefont {Garlatti}, \citenamefont {Salvadori}, \citenamefont {Tacchino}, \citenamefont {Santini}, \citenamefont {Tavernelli}, \citenamefont {Bittl}, \citenamefont {Chiesa}, \citenamefont {Sessoli},\ and\ \citenamefont {Carretta}}]{chiesaAssessingNatureChiralInduced2021}%
  \BibitemOpen
  \bibfield  {author} {\bibinfo {author} {\bibfnamefont {A.}~\bibnamefont {Chiesa}}, \bibinfo {author} {\bibfnamefont {M.}~\bibnamefont {Chizzini}}, \bibinfo {author} {\bibfnamefont {E.}~\bibnamefont {Garlatti}}, \bibinfo {author} {\bibfnamefont {E.}~\bibnamefont {Salvadori}}, \bibinfo {author} {\bibfnamefont {F.}~\bibnamefont {Tacchino}}, \bibinfo {author} {\bibfnamefont {P.}~\bibnamefont {Santini}}, \bibinfo {author} {\bibfnamefont {I.}~\bibnamefont {Tavernelli}}, \bibinfo {author} {\bibfnamefont {R.}~\bibnamefont {Bittl}}, \bibinfo {author} {\bibfnamefont {M.}~\bibnamefont {Chiesa}}, \bibinfo {author} {\bibfnamefont {R.}~\bibnamefont {Sessoli}},\ and\ \bibinfo {author} {\bibfnamefont {S.}~\bibnamefont {Carretta}},\ }\bibfield  {title} {\bibinfo {title} {Assessing the {Nature} of {Chiral}-{Induced} {Spin} {Selectivity} by {Magnetic} {Resonance}},\ }\href {https://doi.org/10.1021/acs.jpclett.1c01447} {\bibfield  {journal} {\bibinfo  {journal} {J. Phys. Chem. Lett.}\ }\textbf {\bibinfo {volume} {12}},\
  \bibinfo {pages} {6341} (\bibinfo {year} {2021})}\BibitemShut {NoStop}%
\bibitem [{\citenamefont {Bloom}\ \emph {et~al.}(2024)\citenamefont {Bloom}, \citenamefont {Paltiel}, \citenamefont {Naaman},\ and\ \citenamefont {Waldeck}}]{bloomChiralInducedSpin2024}%
  \BibitemOpen
  \bibfield  {author} {\bibinfo {author} {\bibfnamefont {B.~P.}\ \bibnamefont {Bloom}}, \bibinfo {author} {\bibfnamefont {Y.}~\bibnamefont {Paltiel}}, \bibinfo {author} {\bibfnamefont {R.}~\bibnamefont {Naaman}},\ and\ \bibinfo {author} {\bibfnamefont {D.~H.}\ \bibnamefont {Waldeck}},\ }\bibfield  {title} {\bibinfo {title} {Chiral {Induced} {Spin} {Selectivity}},\ }\href {https://doi.org/10.1021/acs.chemrev.3c00661} {\bibfield  {journal} {\bibinfo  {journal} {Chem. Rev.}\ }\textbf {\bibinfo {volume} {124}},\ \bibinfo {pages} {1950} (\bibinfo {year} {2024})}\BibitemShut {NoStop}%
\bibitem [{\citenamefont {Tiwari}\ and\ \citenamefont {Poonia}(2023)}]{Poonia2023}%
  \BibitemOpen
  \bibfield  {author} {\bibinfo {author} {\bibfnamefont {Y.}~\bibnamefont {Tiwari}}\ and\ \bibinfo {author} {\bibfnamefont {V.~S.}\ \bibnamefont {Poonia}},\ }\bibfield  {title} {\bibinfo {title} {Quantum coherence enhancement by the chirality-induced spin selectivity effect in the radical-pair mechanism},\ }\href {https://doi.org/10.1103/PhysRevA.107.052406} {\bibfield  {journal} {\bibinfo  {journal} {Phys. Rev. A}\ }\textbf {\bibinfo {volume} {107}},\ \bibinfo {pages} {052406} (\bibinfo {year} {2023})}\BibitemShut {NoStop}%
\bibitem [{\citenamefont {Tiwari}\ and\ \citenamefont {Poonia}(2022)}]{Poonia2022}%
  \BibitemOpen
  \bibfield  {author} {\bibinfo {author} {\bibfnamefont {Y.}~\bibnamefont {Tiwari}}\ and\ \bibinfo {author} {\bibfnamefont {V.~S.}\ \bibnamefont {Poonia}},\ }\bibfield  {title} {\bibinfo {title} {Role of chiral-induced spin selectivity in the radical pair mechanism of avian magnetoreception},\ }\href {https://doi.org/10.1103/PhysRevE.106.064409} {\bibfield  {journal} {\bibinfo  {journal} {Phys. Rev. E}\ }\textbf {\bibinfo {volume} {106}},\ \bibinfo {pages} {064409} (\bibinfo {year} {2022})}\BibitemShut {NoStop}%
\bibitem [{\citenamefont {Fay}\ \emph {et~al.}(2019)\citenamefont {Fay}, \citenamefont {Lindoy},\ and\ \citenamefont {Manolopoulos}}]{fayElectronSpinRelaxation2019}%
  \BibitemOpen
  \bibfield  {author} {\bibinfo {author} {\bibfnamefont {T.~P.}\ \bibnamefont {Fay}}, \bibinfo {author} {\bibfnamefont {L.~P.}\ \bibnamefont {Lindoy}},\ and\ \bibinfo {author} {\bibfnamefont {D.~E.}\ \bibnamefont {Manolopoulos}},\ }\bibfield  {title} {\bibinfo {title} {Electron spin relaxation in radical pairs: {Beyond} the {Redfield} approximation},\ }\href {https://doi.org/10.1063/1.5125752} {\bibfield  {journal} {\bibinfo  {journal} {J. Chem. Phys.}\ }\textbf {\bibinfo {volume} {151}},\ \bibinfo {pages} {154117} (\bibinfo {year} {2019})}\BibitemShut {NoStop}%
\bibitem [{\citenamefont {Zoltowski}\ \emph {et~al.}(2019)\citenamefont {Zoltowski}, \citenamefont {Chelliah}, \citenamefont {Wickramaratne}, \citenamefont {Jarocha}, \citenamefont {Karki}, \citenamefont {Xu}, \citenamefont {Mouritsen}, \citenamefont {Hore}, \citenamefont {Hibbs}, \citenamefont {Green},\ and\ \citenamefont {Takahashi}}]{zoltowskiChemicalStructuralAnalysis2019}%
  \BibitemOpen
  \bibfield  {author} {\bibinfo {author} {\bibfnamefont {B.~D.}\ \bibnamefont {Zoltowski}}, \bibinfo {author} {\bibfnamefont {Y.}~\bibnamefont {Chelliah}}, \bibinfo {author} {\bibfnamefont {A.}~\bibnamefont {Wickramaratne}}, \bibinfo {author} {\bibfnamefont {L.}~\bibnamefont {Jarocha}}, \bibinfo {author} {\bibfnamefont {N.}~\bibnamefont {Karki}}, \bibinfo {author} {\bibfnamefont {W.}~\bibnamefont {Xu}}, \bibinfo {author} {\bibfnamefont {H.}~\bibnamefont {Mouritsen}}, \bibinfo {author} {\bibfnamefont {P.~J.}\ \bibnamefont {Hore}}, \bibinfo {author} {\bibfnamefont {R.~E.}\ \bibnamefont {Hibbs}}, \bibinfo {author} {\bibfnamefont {C.~B.}\ \bibnamefont {Green}},\ and\ \bibinfo {author} {\bibfnamefont {J.~S.}\ \bibnamefont {Takahashi}},\ }\bibfield  {title} {\bibinfo {title} {Chemical and structural analysis of a photoactive vertebrate cryptochrome from pigeon},\ }\href {https://doi.org/10.1073/pnas.1907875116} {\bibfield  {journal} {\bibinfo  {journal} {Proc. Natl. Acad. Sci.}\ }\textbf {\bibinfo {volume} {116}},\
  \bibinfo {pages} {19449} (\bibinfo {year} {2019})}\BibitemShut {NoStop}%
\bibitem [{\citenamefont {Baumgratz}\ \emph {et~al.}(2014)\citenamefont {Baumgratz}, \citenamefont {Cramer},\ and\ \citenamefont {Plenio}}]{Baumgratz2014}%
  \BibitemOpen
  \bibfield  {author} {\bibinfo {author} {\bibfnamefont {T.}~\bibnamefont {Baumgratz}}, \bibinfo {author} {\bibfnamefont {M.}~\bibnamefont {Cramer}},\ and\ \bibinfo {author} {\bibfnamefont {M.~B.}\ \bibnamefont {Plenio}},\ }\bibfield  {title} {\bibinfo {title} {Quantifying coherence},\ }\href {https://doi.org/10.1103/PhysRevLett.113.140401} {\bibfield  {journal} {\bibinfo  {journal} {Phys. Rev. Lett.}\ }\textbf {\bibinfo {volume} {113}},\ \bibinfo {pages} {140401} (\bibinfo {year} {2014})}\BibitemShut {NoStop}%
\bibitem [{\citenamefont {Kaptein}(1972)}]{kapteinChemicallyInducedDynamic1972}%
  \BibitemOpen
  \bibfield  {author} {\bibinfo {author} {\bibfnamefont {R.}~\bibnamefont {Kaptein}},\ }\bibfield  {title} {\bibinfo {title} {Chemically induced dynamic nuclear polarization. {VIII}. {Spin} dynamics and diffusion of radical pairs},\ }\href {https://doi.org/10.1021/ja00773a001} {\bibfield  {journal} {\bibinfo  {journal} {J. Am. Chem. Soc.}\ }\textbf {\bibinfo {volume} {94}},\ \bibinfo {pages} {6251} (\bibinfo {year} {1972})}\BibitemShut {NoStop}%
\bibitem [{\citenamefont {Maeda}\ \emph {et~al.}(2012)\citenamefont {Maeda}, \citenamefont {Robinson}, \citenamefont {Henbest}, \citenamefont {Hogben}, \citenamefont {Biskup}, \citenamefont {Ahmad}, \citenamefont {Schleicher}, \citenamefont {Weber}, \citenamefont {Timmel},\ and\ \citenamefont {Hore}}]{maedaMagneticallySensitiveLightinduced2012}%
  \BibitemOpen
  \bibfield  {author} {\bibinfo {author} {\bibfnamefont {K.}~\bibnamefont {Maeda}}, \bibinfo {author} {\bibfnamefont {A.~J.}\ \bibnamefont {Robinson}}, \bibinfo {author} {\bibfnamefont {K.~B.}\ \bibnamefont {Henbest}}, \bibinfo {author} {\bibfnamefont {H.~J.}\ \bibnamefont {Hogben}}, \bibinfo {author} {\bibfnamefont {T.}~\bibnamefont {Biskup}}, \bibinfo {author} {\bibfnamefont {M.}~\bibnamefont {Ahmad}}, \bibinfo {author} {\bibfnamefont {E.}~\bibnamefont {Schleicher}}, \bibinfo {author} {\bibfnamefont {S.}~\bibnamefont {Weber}}, \bibinfo {author} {\bibfnamefont {C.~R.}\ \bibnamefont {Timmel}},\ and\ \bibinfo {author} {\bibfnamefont {P.~J.}\ \bibnamefont {Hore}},\ }\bibfield  {title} {\bibinfo {title} {Magnetically sensitive light-induced reactions in cryptochrome are consistent with its proposed role as a magnetoreceptor},\ }\href {https://doi.org/10.1073/pnas.1118959109} {\bibfield  {journal} {\bibinfo  {journal} {Proc. Natl. Acad. Sci.}\ }\textbf {\bibinfo {volume} {109}},\ \bibinfo {pages} {4774} (\bibinfo
  {year} {2012})}\BibitemShut {NoStop}%
\bibitem [{\citenamefont {Carmeli}\ \emph {et~al.}(2014)\citenamefont {Carmeli}, \citenamefont {Kumar}, \citenamefont {Heifler}, \citenamefont {Carmeli},\ and\ \citenamefont {Naaman}}]{carmeliSpinSelectivityElectron2014}%
  \BibitemOpen
  \bibfield  {author} {\bibinfo {author} {\bibfnamefont {I.}~\bibnamefont {Carmeli}}, \bibinfo {author} {\bibfnamefont {K.~S.}\ \bibnamefont {Kumar}}, \bibinfo {author} {\bibfnamefont {O.}~\bibnamefont {Heifler}}, \bibinfo {author} {\bibfnamefont {C.}~\bibnamefont {Carmeli}},\ and\ \bibinfo {author} {\bibfnamefont {R.}~\bibnamefont {Naaman}},\ }\bibfield  {title} {\bibinfo {title} {Spin {Selectivity} in {Electron} {Transfer} in {Photosystem} {I}},\ }\href {https://doi.org/10.1002/anie.201404382} {\bibfield  {journal} {\bibinfo  {journal} {Angew. Chem. Int. Ed.}\ }\textbf {\bibinfo {volume} {53}},\ \bibinfo {pages} {8953} (\bibinfo {year} {2014})}\BibitemShut {NoStop}%
\bibitem [{\citenamefont {Eckvahl}\ \emph {et~al.}(2023)\citenamefont {Eckvahl}, \citenamefont {Tcyrulnikov}, \citenamefont {Chiesa}, \citenamefont {Bradley}, \citenamefont {Young}, \citenamefont {Carretta}, \citenamefont {Krzyaniak},\ and\ \citenamefont {Wasielewski}}]{eckvahlDirectObservationChiralityinduced2023}%
  \BibitemOpen
  \bibfield  {author} {\bibinfo {author} {\bibfnamefont {H.~J.}\ \bibnamefont {Eckvahl}}, \bibinfo {author} {\bibfnamefont {N.~A.}\ \bibnamefont {Tcyrulnikov}}, \bibinfo {author} {\bibfnamefont {A.}~\bibnamefont {Chiesa}}, \bibinfo {author} {\bibfnamefont {J.~M.}\ \bibnamefont {Bradley}}, \bibinfo {author} {\bibfnamefont {R.~M.}\ \bibnamefont {Young}}, \bibinfo {author} {\bibfnamefont {S.}~\bibnamefont {Carretta}}, \bibinfo {author} {\bibfnamefont {M.~D.}\ \bibnamefont {Krzyaniak}},\ and\ \bibinfo {author} {\bibfnamefont {M.~R.}\ \bibnamefont {Wasielewski}},\ }\bibfield  {title} {\bibinfo {title} {Direct observation of chirality-induced spin selectivity in electron donor–acceptor molecules},\ }\href {https://doi.org/10.1126/science.adj5328} {\bibfield  {journal} {\bibinfo  {journal} {Science}\ }\textbf {\bibinfo {volume} {382}},\ \bibinfo {pages} {197} (\bibinfo {year} {2023})}\BibitemShut {NoStop}%
\bibitem [{\citenamefont {Katsoprinakis}\ \emph {et~al.}(2010)\citenamefont {Katsoprinakis}, \citenamefont {Dellis},\ and\ \citenamefont {Kominis}}]{katsoprinakisCoherentTripletExcitation2010}%
  \BibitemOpen
  \bibfield  {author} {\bibinfo {author} {\bibfnamefont {G.~E.}\ \bibnamefont {Katsoprinakis}}, \bibinfo {author} {\bibfnamefont {A.~T.}\ \bibnamefont {Dellis}},\ and\ \bibinfo {author} {\bibfnamefont {I.~K.}\ \bibnamefont {Kominis}},\ }\bibfield  {title} {\bibinfo {title} {Coherent triplet excitation suppresses the heading error of the avian compass},\ }\href {https://doi.org/10.1088/1367-2630/12/8/085016} {\bibfield  {journal} {\bibinfo  {journal} {New J. Phys.}\ }\textbf {\bibinfo {volume} {12}},\ \bibinfo {pages} {085016} (\bibinfo {year} {2010})}\BibitemShut {NoStop}%
\bibitem [{\citenamefont {Lin}\ \emph {et~al.}(2025)\citenamefont {Lin}, \citenamefont {Tsuji}, \citenamefont {Bruzzese}, \citenamefont {Chen}, \citenamefont {Vrionides}, \citenamefont {Jian}, \citenamefont {Kittur}, \citenamefont {Fay},\ and\ \citenamefont {Mani}}]{linMolecularEngineeringEmissive2025}%
  \BibitemOpen
  \bibfield  {author} {\bibinfo {author} {\bibfnamefont {N.}~\bibnamefont {Lin}}, \bibinfo {author} {\bibfnamefont {M.}~\bibnamefont {Tsuji}}, \bibinfo {author} {\bibfnamefont {I.}~\bibnamefont {Bruzzese}}, \bibinfo {author} {\bibfnamefont {A.}~\bibnamefont {Chen}}, \bibinfo {author} {\bibfnamefont {M.}~\bibnamefont {Vrionides}}, \bibinfo {author} {\bibfnamefont {N.}~\bibnamefont {Jian}}, \bibinfo {author} {\bibfnamefont {F.}~\bibnamefont {Kittur}}, \bibinfo {author} {\bibfnamefont {T.~P.}\ \bibnamefont {Fay}},\ and\ \bibinfo {author} {\bibfnamefont {T.}~\bibnamefont {Mani}},\ }\bibfield  {title} {\bibinfo {title} {Molecular {Engineering} of {Emissive} {Molecular} {Qubits} {Based} on {Spin}-{Correlated} {Radical} {Pairs}},\ }\bibfield  {journal} {\bibinfo  {journal} {J. Am. Chem. Soc.}\ }\href {https://doi.org/10.1021/jacs.4c16164} {10.1021/jacs.4c16164} (\bibinfo {year} {2025})\BibitemShut {NoStop}%
\bibitem [{\citenamefont {Mani}(2022)}]{maniMolecularQubitsBased2022}%
  \BibitemOpen
  \bibfield  {author} {\bibinfo {author} {\bibfnamefont {T.}~\bibnamefont {Mani}},\ }\bibfield  {title} {\bibinfo {title} {Molecular qubits based on photogenerated spin-correlated radical pairs for quantum sensing},\ }\href {https://doi.org/10.1063/5.0084072} {\bibfield  {journal} {\bibinfo  {journal} {Chem. Phys. Rev.}\ }\textbf {\bibinfo {volume} {3}},\ \bibinfo {pages} {021301} (\bibinfo {year} {2022})}\BibitemShut {NoStop}%
\bibitem [{\citenamefont {Chowdhury}\ \emph {et~al.}(2024)\citenamefont {Chowdhury}, \citenamefont {Denton}, \citenamefont {Bonser},\ and\ \citenamefont {Kattnig}}]{chowdhuryQuantumControlRadicalPair2024}%
  \BibitemOpen
  \bibfield  {author} {\bibinfo {author} {\bibfnamefont {F.~T.}\ \bibnamefont {Chowdhury}}, \bibinfo {author} {\bibfnamefont {M.~C.}\ \bibnamefont {Denton}}, \bibinfo {author} {\bibfnamefont {D.~C.}\ \bibnamefont {Bonser}},\ and\ \bibinfo {author} {\bibfnamefont {D.~R.}\ \bibnamefont {Kattnig}},\ }\bibfield  {title} {\bibinfo {title} {Quantum {Control} of {Radical}-{Pair} {Dynamics} beyond {Time}-{Local} {Optimization}},\ }\href {https://doi.org/10.1103/PRXQuantum.5.020303} {\bibfield  {journal} {\bibinfo  {journal} {PRX Quantum}\ }\textbf {\bibinfo {volume} {5}},\ \bibinfo {pages} {020303} (\bibinfo {year} {2024})}\BibitemShut {NoStop}%
\end{thebibliography}%

\end{document}